\documentclass{jfm}

\usepackage{graphicx}
\graphicspath{{/Figure/}}
\usepackage[justification=justified,format=plain, width=\textwidth]{caption}
\usepackage{epstopdf}
\usepackage{newtxtext}
\usepackage{newtxmath}
\usepackage{slashbox}
\usepackage{natbib}
\usepackage{hyperref}
\hypersetup{
    colorlinks   = true, %Colours links instead of ugly boxes
    urlcolor     = blue, %Colour for external hyperlinks
    linkcolor    = blue, %Colour of internal links
    citecolor   = blue %Colour of citations
}

\newcommand*{\figref}[2][]{%
  \hyperref[{#2}]{%
    figures~\ref*{#2}%
    \ifx\\#1\\%
    \else
      \,#1%
    \fi
  }%
}

\newcommand*{\Figref}[2][]{%
  \hyperref[{#2}]{%
    Figure~\ref*{#2}%
    \ifx\\#1\\%
    \else
      \,#1%
    \fi
  }%
}

\newcommand*{\Figsref}[2][]{%
  \hyperref[{#2}]{%
    Figures~\ref*{#2}%
    \ifx\\#1\\%
    \else
      \,#1%
    \fi
  }%
}

\usepackage{cleveref}

\usepackage{mathtools, nccmath}
\usepackage{epstopdf}
\usepackage{graphicx,stfloats}
\usepackage{setspace,lipsum}
\usepackage{amsmath}
\usepackage[dvipsnames]{xcolor}
\usepackage{epstopdf, epsfig}
\usepackage{natbib}
\usepackage{mathrsfs}
\usepackage{array}
\usepackage[scr=rsfs]{mathalpha}

\newcommand{\RomanNumeralCaps}[1]
\linenumbers

\title{Coupled thermoacoustic interactions in hydrogen enriched lean combustion}

\author{Abhishek Kushwaha\aff{1,2}
  \corresp{\email{abhikushwaha8090@gmail.com}},
  Amitesh Roy\aff{3},
  Ianko Chterev\aff{4},
  Isaac Boxx\aff{5},
  R. I. Sujith\aff{1,2}}

\affiliation{\aff{1}Department of Aerospace Engineering, IIT Madras, Chennai, Tamil Nadu, 600036, India
\aff{2} Centre of Excellence for Studying Critical Transition in Complex Systems, IIT Madras, Chennai, Tamil Nadu, 600036, India
\aff{3}Institute for Aerospace Studies, University of Toronto, Ontario M3H5T6, Canada
\aff{4}Institute for Combustion Technology,
German Aerospace Centre (DLR),
Stuttgart 70569, Germany
 \aff{5}Chair of Optical Diagnostics for Energy, Process and Chemical Engineering. RWTH Aachen University,
 Aachen 52062, Germany}

\begin{document}
\maketitle

\large \textbf{Abstract}

\begin{abstract}

In this paper, we present a framework to study the synchronization of flow velocity with acoustic pressure and heat-release rate in technically-premixed swirl flames. The framework uses the extended proper orthogonal decomposition to identify regions of the velocity field where velocity and heat release fluctuations are highly correlated. We apply this framework to study coupled interactions associated with period-1 and period-2 type thermoacoustic instability in a technically-premixed, swirl stabilized gas turbine model combustor operated with hydrogen-enriched natural gas. We find the structures in flame surface and heat release rate that are correlated to the dominant coherent structures of the flow field using extended POD. We observe that the correlated structures in the flow velocity, flame surface and heat release rate fields share the same spatial regions during thermoacoustic instability with period-1 oscillations. In the case of period-2 oscillations, the structures from flame surface and heat release rate field are strongly correlated. However, these structures contribute less to the coherent structures of the flow field. Using the temporal coefficients of the dominant POD modes of the flow velocity field, we also observed 1:1 and 2:1 frequency locking behaviour among the time series of acoustic pressure, heat release rate and the temporal coefficients of the first two dominating POD modes of velocity field during the state of period-1 and period-2 oscillations, respectively. These frequency-locked states, which indicate the underlying phase-synchronization states, are then correlated with coherent structures in the flow velocity field.

\end{abstract}

%\begin{keywords}

%\end{keywords}

%{\bf MSC Codes }  {\it(Optional)} Please enter your MSC Codes here

\section{Introduction}
%\setstretch{2}

Reducing global greenhouse emissions is essential to ensure the habitability of our planet. One of the major sources of these emissions comes from the combustion of hydrocarbon fuels which continues to be the primary source of power generation in modern gas turbines utilized in aviation and thermal power plants. Lean combustion with the addition of hydrogen has become a significant alternative for enhancing lean combustion limits, providing high energy, and lowering carbon emissions. \citep{janus1997effects, schefer2003hydrogen, birol2019future, hord1978hydrogen}. 

Despite evident advantages, the effect of hydrogen enrichment on the thermoacoustic stability of turbulent combustors is not well-understood and remains an active topic of research. Turbulent lean premixed flames are particularly sensitive to acoustic fluctuations. Under suitable conditions, the positive feedback among the acoustic field, turbulence and the flame leads to thermoacoustic instability which results in large-amplitude pressure fluctuations \citep{lieuwen2021unsteady, sujith2021thermoacoustic}. Thermoacoustic instability (TAI) develops when heat release rate and pressure oscillations occur in-phase and the acoustic driving is greater than acoustic damping, leading to the energy increment to the acoustic field of the combustor \citep{rayleigh1878explanation,chu1965energy,putnam1971combustion}. The amplitude of pressure oscillations increases nonlinearities in the acoustic damping and the acoustic driving \citep{lieuwen2005combustion,sujith2021thermoacoustic}. The balance between the acoustic losses and growth leads to limit cycle oscillations \citep{dowling1997nonlinear}. The state of TAI can be detrimental to components of the engine, imparting structural damages, resulting in unscheduled shutdowns and mission failures \citep{juniper2018sensitivity}.

Large-scale coherent structures emerge during the onset of combustion instability in combustion systems \citep{poinsot1987vortex, sterling1987longitudinal, george2018pattern}. The mutual synchronization of the reactive flow field with pressure oscillations is associated with the shift from stable combustion operation to the state of TAI \citep{mondal2017synchronous, pawar2017thermoacoustic, singh2022mean}. The goal of this study is to present a new framework based on the extended POD for quantifying the synchronization of coherence structures in the flow with heat release rate and pressure oscillations in a swirl-stabilized combustor. To accomplish this, we quantify the mutual synchronization of these coherent flow structures with pressure oscillations during various dynamical states attained at various levels of hydrogen enrichment.

\subsection{Dynamics of swirling flows and flames} 

Swirling flames are used extensively in aircraft and gas turbine engines as they are one of the simplest and robust methods of stabilizing turbulent flame \citep{gupta1984swirl, hoffmann1994development,reddy2006swirler,freitag2007investigation, candel2014dynamics}. The use of swirl flow leads to better air-fuel mixing \citep{syred2014effect}, reduction in fuel consumption and decrement in exhaust pollutant emission \citep{rashwan2016review}. 

Intense swirling flow gives rise to a central recirculation zone or a ‘vortex breakdown bubble’ (VBB) \citep{harvey1962some}, which provides hot gases to the flame at the base, improving the static stability of the flame. The flow returns along the flow center-line when the swirl number exceeds a critical threshold, forming an inner recirculation zone (IRZ) through helical instability \citep{gupta1984swirl,choi2007numerical}. The axial flux of azimuthal momentum divided by the axial flux of axial momentum is known as the swirl number. At sufficiently intense swirling conditions, precessing vortex cores (PVCs) form through helicoidal instabilities \citep{liang2005experimental,gallaire2006spiral, qadri2013structural}. On the other hand, the outer recirculation zone (ORZ) is formed when swirling flows encounter sudden expansion, leading to the flow recirculation along the dump plane. Turbulent flames can stabilize either in the central or IRZ or in the ORZ or in both zones depending upon the overall geometry of the combustor \citep{gupta1984swirl}. Further, the hydrodynamic stability is altered significantly due to the presence of density stratification in non-isothermal conditions of turbulent flames. Indeed, density stratification has been shown to suppress the formation of PVCs \citep{oberleithner2013nonuniform}. Thus, the overall flow behavior is strongly affected by the global stability of swirling flows and flames \citep{oberleithner2011three, stohr2011phase}.

\subsection{Effect of hydrogen enrichment on swirling turbulent flames}

Hydrogen enrichment can significantly alter the characteristics of turbulent flames. Comparing the size of the reaction zone to a pure methane-air flame stabilized by swirling, \cite{wicksall2005interaction} noticed that the addition of hydrogen to methane makes the flame more resilient. The increased concentration of OH, H, and O radicals in the ensuing $\textrm{H}_2$-enriched flame directly contributes to the increase of flame stability limits \citep{schefer2003hydrogen} in premixed turbulent combustors \citep{cozzi2006behavior}. One key feature of $\textrm{H}_2$-enrichment is the much higher burning velocity of the enriched methane mixture relative to pure methane due to the fast reaction rate of hydrogen observed in both numerical simulations \citep{hawkes2004direct, nam2019numerical, guo2020effect,xia2022numerical} and experiments \citep{mandilas2007effects, strakey2007investigation, nakahara2008study, zhang2020effect, pignatelli2022pilot}. 

Higher reaction rates also manifest in much larger increment in flame surface density upon $\textrm{H}_2$-enrichment  \citep{halter2007characterization, guo2010burning} and enhancement of flame wrinkling due to interactions of small-scales of turbulence with the flame front \citep{emadi2012flame}. Further, addition of hydrogen alters the manner in which flame interacts with turbulence and pressure fluctuations, thereby leading to increased rate of reaction of hydrogen blended methane \citep{kim2009hydrogen}. The addition of $textrmH_2$ enhances flame anchoring and lean combustion limits by strengthening the flame's resistance to strain brought on by turbulence \citep{schefer2002combustion, zhang2011strain}. 

%\subsection{flame shape and thermoacoustic behavior}

As such, H$_2$ enrichment in methane increases the flame surface area and the flame wrinkling and also contributes to the state of TAI by increasing the heat release rate fluctuations \citep{zhang2019experimental}. Moreover, the overall change in the flame shape at various levels of $\textrm{H}_2$-enrichment is also a strong contributor to TAI while affecting pollutant emissions \citep{schmitt2007large}. \cite{tuncer2009dynamics} reported that a shift in the flame center of mass towards the inlet due to an increase in the burning velocity with hydrogen enrichment, controls the shape of the flame. Similarly, swirling methane-air flame under different levels of H$_2$ enrichment undergoes drastic change from columnar, V-shape to M-shape flame \citep{davis2013effects, taamallah2015thermo, shanbhogue2016flame, chterevflame}. In fact, these effects persist at elevated pressures at various levels of H$_2$ enrichment \citep{chterev2021effect} which determine the exact flame shape depending upon the stability of the swirling flow. \cite{garcia2014effect} found that a V-flame transitions to an M-flame whenever the swirling flow features an outer reaction zone. 

Using linear stability analysis, \cite{oberleithner2015formation} showed that V-shape flames possess strong radial gradients, thus suppressing the formation of precessing vortex cores (PVC) and the flame predominantly anchors along the shear layer. In contrast, M-flames anchor along the outer recirculation zones and possess smooth density gradients. These flames also exhibit PVCs. However, the role of PVCs in driving thermoacoustic instabilities have not yet been resolved satisfactorily \citep{candel2014dynamics}. While initially thought to be a contributor of thermoacoustic instability \citep{gupta1984swirl, huang2009dynamics}, in many cases PVCs have been found to have limited impact on the thermoacoustic characteristics of axisymmetric swirl combustors \citep{boxx2010temporally, moeck2012nonlinear, oberleithner2013nonuniform}. This is due to PVC-induced anti-symmetric flame perturbations, which have no impact on oscillations in the global heat release rate \citep{candel2014dynamics}.    

In the specific context of thermoacoustic instability in swirl combustors, addition of $\textrm{H}_2$ can shift the state to TAI from combustion noise \citep{janus1997effects}. Here, the term "combustion noise" refers to a steady operating condition characterized by low amplitude aperiodic acoustic oscillations with broadband spectra \citep{candel2009flame} and fractal characteristics \citep{nair2014multifractality}. The addition of hydrogen shifts the occurrence of the state of TAI \citep{figura2007effects} to lower equivalence ratios with the reduction in dynamic pressure amplitude. These characteristics were ascribed to the hydrogen present in the combination of $textrm CH_4$ and $textrm H_2$, which caused a higher rate of reaction and a shorter convective time scale. \cite{hong2013phase} reported a varying frequency and flame response when hydrogen was varied as a result of change in the relative phase between heat release rate and pressure fluctuations. In addition to seeing several developments related to flame transfer function, \cite{aesoy2020scaling} showed a reduction in the phase difference between heat release rate and pressure for hydrogen-enriched fuel. \cite{lee2020combustion} separately investigated the dynamics of a mesoscale burner for the combustion of methane and hydrogen and reported that the dynamics is connected with higher eigenmodes for pure hydrogen operation. Similarly, many research studies show different effects of hydrogen-enrichment on the combustion dynamics \citep{taamallah2015thermo,shanbhogue2016flame,zhang2019experimental,chterevflame}.

\subsection{Modal decomposition of thermoacoustic systems}

Coherent structures are intimately related to the overall thermoacoustic behaviour of turbulent combustors \citep{poinsot1987vortex,sterling1987longitudinal,schadow1992combustion} as these structures dictate the behaviour of the flow. Using techniques such as spark-Schlieren imaging and C$_2$ radiation mapping, these studies tracked the formation, growth and decay of coherent structures. These structures affect the mixing process of the unburnt fuel-air mixture with burnt hot gases. The mutual synchronization of the pressure fluctuations with the periodic shedding of coherent structures are one of the key mechanisms underlying the emergence of the state of TAI \citep{poinsot1987vortex, pawar2017thermoacoustic, george2018pattern}. 

One of the most significant techniques for analysing coherent patterns in turbulent flow is the proper orthogonal decomposition (POD) approach \citep{lumley1967structure, sirovich1987turbulence}. POD method reduces turbulent flow fields in terms of orthogonal modes which are obtained by optimizing the $\boldsymbol{\mathit{L}}^2$-norm or the kinetic energy of the velocity field. The POD modes are orthogonal and characterize the most energy containing coherent structures. Fluid dynamical systems have strongly coupled subsystems and to understand the effect of coherent structures in the flow on other correlated variables, \cite{boree2003extended} introduced the extended proper orthogonal decomposition (EPOD). The method is used to project other correlated variables such as concentration, temperature, etc., on the POD modes of the velocity field and obtain analogous flow structures in such correlated variables.

In the context of thermoacoustics, the relationship between coherent structures and pressure variations has been investigated with POD. For instance, \cite{sui2017experimental} used the POD method to understand the behavior of acoustic fluctuations of different eigenmodes in space using the reconstructed 2-D pressure field of a Rijke tube. They reported that the heating source mitigates the pressure fluctuations. They also showed 1:1 and 2:1 coupling among different filtered modal coefficients using the traditional Lissajous patterns (axisymmetric patterns). The Lissajous patterns are the representation method for a pair of signals by which the frequency locking behaviour can be provided in terms of the ratio between the dominating frequencies of the pair of the signals \citep{greenslade1993all}. \cite{boxx2010temporally} used POD to study PVC and showed that the dominant vortices in the most energetic mode are rotating in the counter-clockwise direction. However, the second mode has the dominating vortices rotating clockwise. With the help of POD modes, they reported that the first two dominating modes represent a helical PVC and the third modes is associated with the thermoacoustic pulsation in the axial direction. \cite{duwig2010extended} used EPOD to study the correlation among the modes of the reacting flow field and the unsteady flame using PLIF data for the acoustically excited flame. Similarly, EPOD has also been used to determine the effect of coherent structures on correlated modes of related variables such as pressure, temperature, scalar concentrations (OH* and CH* intensity fields) in various thermoacoustic systems \citep{sieber2017advanced,  wang2019proper, lohrasbi2021modification}.

\subsection{Contributions of the present study}
In this study, we aim to provide a framework to understand the synchronization of the flow velocity field with heat release rate and acoustic pressure and to clarify the manner in which the addition of $\textrm{H}_2$ affects coherent flow structures of a highly turbulent swirling flame, influence heat release rate fluctuations and modify the thermoacoustic behavior in a partially premixed, swirl-stabilized flame. To characterize the coupled interaction, we use POD to obtain the dominant modes of the measured velocity field. With the help of dominating POD modes, we unravel their effect on the heat release rate fluctuations through the use of EPOD modes. In this study, we are referring the $x$-component of the velocity ($u_x$), acquired in the Cartesian frame of reference as transverse component and the $y$-component of the velocity ($u_y$) as the axial component of the flow field. This also enables us to correlate how the thermoacoustic behaviour is affected by the addition of $\textrm{H}_2$. We compare the behavior of such extended POD modes during the state of period-1 and period-2 thermoacoustic instability \citep{kushwaha2021dynamical} and contrast their characteristics with the baseline case of combustion noise. By analyzing the spatial EPOD modes simultaneously, we relate the behavior of the dominant modes of the swirling flow, the flame structure and the heat release rate field with the self-excited acoustic mode of the combustor. In this manner, we delineate the overall synchronization behaviour of the swirling flames during various dynamical states and relate them to the dominant modes or the coherent structure of the swirling flow.

The rest of the paper is organized as follows. In §\ref{experimental setup}, the details of the experimental set-up and the imaging techniques used are provided. In §\ref{methodology}, we discuss the methodology used for characterizing the coupled interactions and check the existence of the chaos. This is followed by results and discussion in §\ref{section:Results and discussion}. We summarize our major findings in §\ref{sec:Conclusion}.

\section{Experimental setup}
\label{experimental setup}

\begin{figure}

  \centerline{\includegraphics[scale = 0.46]{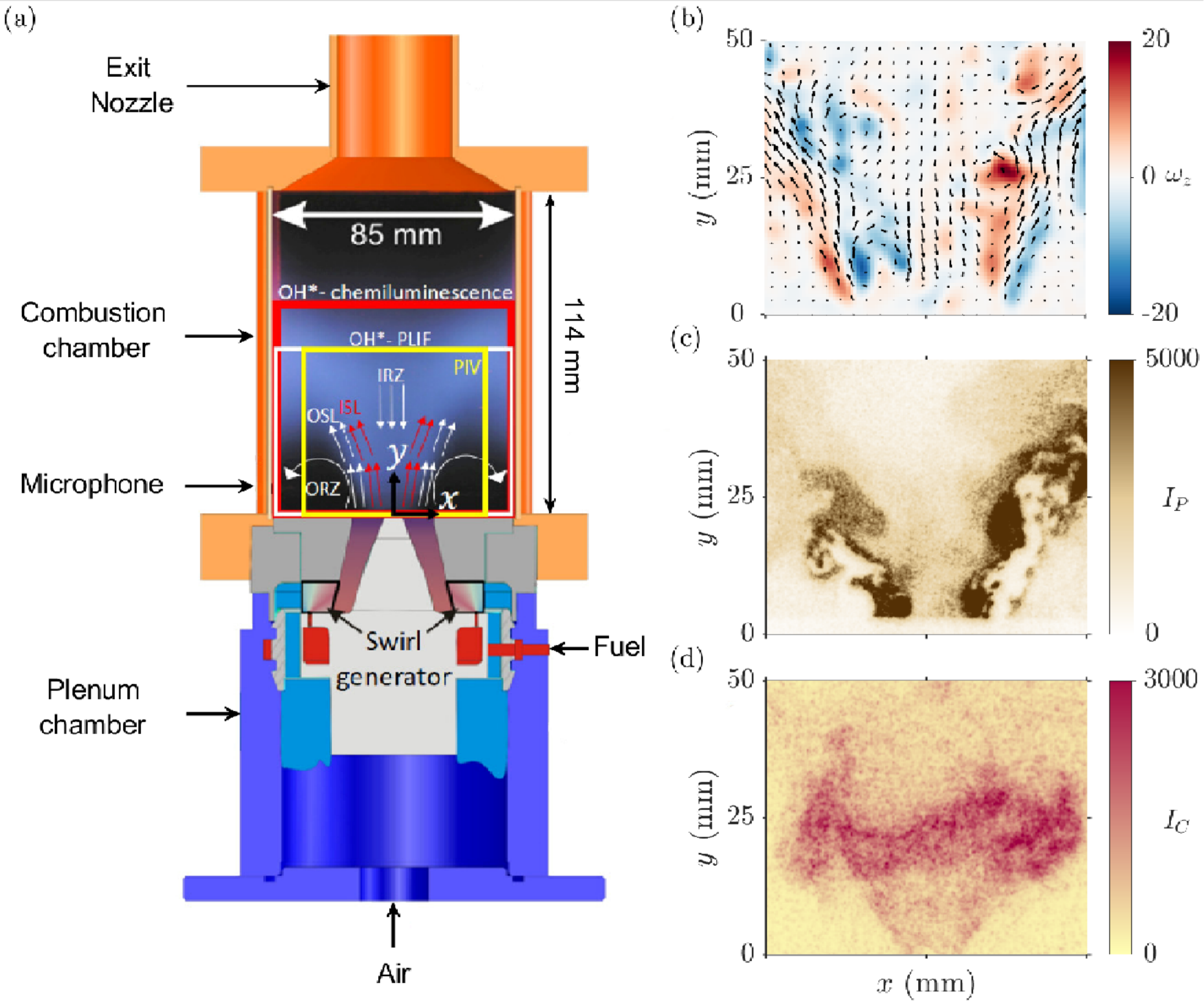}}% Images in 100% size
  \caption{(a) Schematic of the swirl-stabilized PRECCINSTA combustor. Various features of the swirling flow are identified: OSL - outer shear layer, ISL - inner shear layer, ORZ - outer recirculation zone and IRZ - inner recirculation zone. The domain of particle image velocimetry (PIV), OH*-chemiluminescence and planar laser-induced fluorescence (PLIF) imaging are also indicated \citep{kushwaha2021dynamical}. Instantaneous (b) vorticity field, (c) OH-PLIF intensity showing the flame area, and (d) OH*-chemiluminescence field showing the heat release rate distribution are also shown for reference.}
\label{fig1}
\end{figure}

 The experiments were performed in a technically premixed gas turbine model combustor (PRECCINSTA configuration) shown in figure \ref{fig1}. After passing through a swirl generator embedded with 12 radial swirl channels, air enters the cylindrical plenum. Fuel (methane with variable quantities of hydrogen addition) is injected into each swirl channel through a 1 mm orifice. The reactant mixture then passes through a conical nozzle with an exit diameter ($D= 27.85$ mm) and reaches the combustion chamber. Within the nozzle is a conical centerbody with a rounded tip. The chamber has a height of 114 mm and a square cross-section area of 85 $\times$ 85 mm$^2$. The combustor has quartz windows and provide optical access for performing various flow diagnostics. For all the experiments reported here, the combustor is maintained at the same initial conditions. 

Experiments involving the dynamics of the state of TAI of $\textrm{CH}_4$-air flame at various levels of $\textrm{H}_2$ enrichment are performed while keeping the global equivalence ratio constant. This is ensured by simultaneously decreasing $\dot{m}_{\rm CH_4}$ and increasing $\dot{m}_{\rm H_2}$. The dynamical states, analysed in this paper, have the operating conditions, indicated in table \ref{table1}. There is also a slight variation from 22.17 kW to 25.31 kW in the thermal power $P_\textrm{th}$ when the level of $\textrm{H}_2$ is varied. Based on the nozzle's exit diameter, the nominal Reynolds number is in the order of $2 x 10^4$. A more comprehensive explanation of the experimental setup, instrumentation, equipment sensitivity and uncertainty, and optical diagnostics are presented in \S~S1 of supplemental materials.

%Various measurements were performed simultaneously and in a phase-locked manner, to understand the dynamical state of the system. Microphone, positioned at 20 mm from the dump plane of the combustor, was used to record acoustic pressure fluctuations data at a sampling frequency of 100 kHz. Stereoscopic particle image velocimetry (sPIV) was used to measure the flow characteristics inside the combustor. PIV images were acquired at a resolution of 640 $\times$ 800 pixels for each operational state at a resolution of 0.08 mm/pixel and sampling frequency of 10 kHz. The PIV measurement domain spans a region of 65 $\times$ 50 mm$^2$, ($-32 < x < 32$ mm and $0 < y < 50$ mm in figure \ref{fig1}). The heat release rate distribution is determined from images of line of sight integrated chemiluminescence from the self-excited OH* radical (in the bandwidth $300–325$ nm) in the flow field. The chemiluminescence imaging was acquired over 512 $\times$ 512 pixels at sampling frequency of 10 kHz. Finally, planar laser induced fluorescence (PLIF) of OH radicals at a bandwidth of $300-325$ nm was performed to obtain the flame edge fluctuations. PLIF measurements were acquired over a 768 $\times$ 768 pixels at a resolution of 0.115 mm/pixel and sampling frequency of 10 kHz. 

\begin{table}
  \begin{center}
  \begin{tabular}{lcccccc}
      Sr. No.  & $\textrm{H}_2 (\%)$  & $Re (\times 10^4) $ & $\phi_g$ & $P_\textrm{th}$ (kW) & $p^\prime_\textrm{rms}$ (kPa) & Dynamical state \\[3pt]%& $S$ 
      \hline
       1 & 0  & 2.41 & 0.65 & 25.31 &0.09 & Chaotic \\%& \textbf{0.76}\\
       %2  & 20  & 2.37 & 0.65 & 25.28 & 0.06 & Chaotic \\ %& 0.84\\
       %3  & 40  & 2.32 & 0.65 & 25.21 & 0.07 & Chaotic \\%& 1.33\\
       %4  & 0   & 2.78 & 0.65 & 22.17 & 0.39 & P2 LCO \\%& 0.62\\
       %5  & 10  & 2.76 & 0.65 & 22.21 & 0.81 & P2 LCO \\%&0.80\\
       2  & 20  & 2.74 & 0.65 & 22.26 & 0.91 & P2 LCO \\%& \textbf{0.89}\\
      % 7  & 30  & 2.71 & 0.65 & 22.33 & 0.71 & P2 LCO\\%& 0.85\\
      % 8  & 40  & 2.68 & 0.65 & 22.40 & 0.34 & Intermittency \\%& 0.88\\
       3  & 50 & 2.64 & 0.65 & 22.50 & 0.71 & P1 LCO \\%& \textbf{1.19}\\
  \end{tabular}
  \caption{The values of different operating parameters during the experiments. Keys: $\textrm{H}_2 (\%)$ - Volumetric percentage of $\textrm{H}_2$ in the fuel; $Re$ - Nominal Reynolds number; $\phi_g$ - Global equivalence ratio; $P_\textrm{th}$ - Thermal power rating (kW) and dynamical states. The experimental cases analysed here is highlighted. } 
  \label{table1}
  \end{center}
\end{table}

%Figure \ref{fig1}b shows the velocity field obtained from the stereo PIV measurements for a reference case. Figures \ref{fig1}c,d shows the tomographic section of the flame surface and heat release rate distribution obtained using OH-PLIF and chemiluminescence measurements, respectively. Figure \ref{fig1}d of manuscript emphasizes the 3-D flame structure and the associated heat release rate distribution.

\section{Method of analysis}
\label{methodology}

\subsection{Extended proper orthogonal decomposition for obtaining correlated flow structures}
\label{EPOD}

Proper orthogonal decomposition (POD) is a data-driven technique to reduce large-scale, high-dimensional processes or data-sets into low-dimensional deterministic modes \citep{berkooz1993proper,oberleithner2015formation,taira2017modal}. POD decomposes snapshots of space and time-dependent observables into orthogonal modes, ranks them according to their energy and obtains the spatial structure of the corresponding modes. Here, we use the snapshot POD method suggested by \cite{sirovich1987turbulence}. It has a spatial average operator to extract coherent structures from the time-resolved snapshots of fields such as velocity, pressure, temperature and concentration. The method uses $\boldsymbol{\mathit{L}}^2$-norm for classifying the modes, which is equivalent to the fluctuating kinetic energy for the velocity field and variance for other fields.

The velocity fluctuations $(\mathbf{u}^\prime)$ can be decomposed in terms of the spatial modes $(\boldsymbol{\psi}_i)$ and temporal modes $(\boldsymbol{a}_i)$ in the following manner:
\begin{equation}
\mathbf{u}^{\prime} (x,y,t) := \mathbf{u}(x,y,t) - \bar{\mathbf{u}} (x,y) := \sum^N_{i=1} a_i (t) \ \boldsymbol{\Psi_i} (x,y),
\end{equation}
where, $\bar{\mathbf{u}}(x,y)$ refers to time average of the instantaneous velocity field $\mathbf{u}(x,y,t)=u_x \hat{i}+u_y\hat{j}+u_z\hat{k}$ with a resolution of $m \times n$. Depending upon the component of velocity of interest, we reshape each image as a row and form a two-dimensional matrix that has a size of $N \times (m \times n)$ \citep{taira2017modal}, where $N$ is the total number of images. Each snapshot is considered to be an element of the square-integrable vector field $\boldsymbol{\mathit{L}}^2 (\xi)$ specified in the Hilbert space. The class of square-integrable functions is unique for compatibility with an inner product, which allows us to define conditions of orthogonality. The inner product of two vector $\boldsymbol{\alpha}$ and $\boldsymbol{\beta}$ fields in the Hilbert space is defined as 
\begin{equation}
    (\boldsymbol{\alpha,\beta})_{\xi} \coloneqq \int_{\xi} \boldsymbol{\alpha} \cdot \boldsymbol{\beta} \ d\boldsymbol{x},
\end{equation}
where, in a spatial domain $\xi\subset \rm I\!R^3$, $\boldsymbol{x}$ is a point and the corresponding norm $||\boldsymbol{\alpha}||_{\xi}$ is given by
\begin{equation}
    ||\boldsymbol{\alpha}||_{\xi} \coloneqq \sqrt{(\boldsymbol{\alpha} , \boldsymbol{\alpha})_{\xi}}.
\end{equation}
Usually, Compared to the number of snapshots, the number of spatial points is significantly bigger. So, we quantify the relation between individual snapshots by formulating the correlation matrix \textbf{R}$_{(N \times N)}$ 
\begin{equation}
R_{i,j} \coloneqq \frac{1}{N} \left[\mathbf{u}^{\prime} \left(\mathbf{x},t_i\right), \mathbf{u}^{\prime} \left(\mathbf{x},t_j\right)^T\right].
\label{Eq-CorrelationMatrix}
\end{equation}
Here, each snapshot is stored in $\mathbf{x} = x_i$ after reshaping from size $(m$ × $n)$, and $T$ indicates the transpose of the matrix.

Using eigenvalue decomposition, we calculate the eigenvectors and eigenvalues of the correlation matrix \textbf{R},
\begin{equation}
\centering
\textbf{R} a_i = \lambda_i  a_i.
\end{equation}
\textbf{R} has real and non-negative eigenvalues $\lambda_i \geq 0$. We assume the eigenvalues are arranged in decreasing order without losing generality, i.e. $\lambda_1 \geq \lambda_2 \geq ..... \geq \lambda_N.$ The eigenvectors $\textbf{a}_i$ are called the temporal coefficients of the velocity field and follow the orthogonality condition   
\begin{equation}
\frac{1}{N} \sum^N_{k=1} a_i (t_k) a_j (t_k) = \lambda_i \delta_{ij},
\end{equation}
where, $\delta_{ij}$ is the Kronecker Delta function.  

The spatial modes are determined as follows using the projection of the snapshots on the temporal coefficients,
\begin{equation}
\boldsymbol{\Psi_i} (\mathbf{x}) := \frac{1}{N\lambda_i} \sum^N_{j=1} a_i(t_j) \mathbf{u}^{\prime} (\mathbf{x},t_j).
\label{Eq-SpatialPODModes}
\end{equation}
The spatial modes are orthogonal by construction, viz, $(\mathbf{\Psi}_p,\mathbf{\Psi}_q)_\xi=\delta_{pq}$. After reshaping $\boldsymbol{\Psi_i} (\mathbf{x})$, we can get spatial modes $\boldsymbol{\Psi_i} (x,y)$ having $m$ and $n$ as rows and columns, respectively.

Additionally, The correlation between the flow structures and structures in other quantities such as concentration, temperature, or pressure, can be identified using the POD method. This can be achieved by the extended POD analysis \citep{boree2003extended}. The velocity field $u (\mathbf{x},t)$ is measured concurrently with a collection of intensity fields $I (\mathbf{x},t)$. $I (\mathbf{x},t)$ is first decomposed into the average $\bar{I} (\mathbf{x},t)$ and a fluctuating part $I^{\prime} (\mathbf{x},t)$
\begin{equation}
I (\mathbf{x},t)= \bar{I} (\mathbf{x}) + I^{\prime} (\mathbf{x},t). 
\end{equation}
Similar to the spatial modes $\Psi (\mathbf{x})$ of velocity, a set of extended POD modes $\boldsymbol{\Omega}_i$({\textbf{x}}) can be defined as 
\begin{equation}
\boldsymbol{\Omega_i} (\mathbf{x}) = \frac{1}{N\lambda_i} \sum^N_{j=1} \textbf{a}_i(t_j) I^{\prime} (\mathbf{x},t_j),
\end{equation}
where $\boldsymbol{a_i} (t_j)$ are the temporal coefficients of the POD modes of velocity fields. In this manner, we correlate coherent structures observed in the swirling flow with the extended modes in the OH-PLIF ($I_P$) and OH*-chemiluminescence fields ($I_C$).

\subsection{Temporal synchronization of coherent structures}
\label{Temporal synchronization}

The POD and extended POD analysis quantifies the spatial extent of flow and flame structures. The temporal interactions of these spatially extended modes are equally important, and we quantify these interactions based on their synchronization characteristics. This is achieved by time series analysis of the POD temporal coefficients along with pressure and global heat release rate fluctuations. 

We compute the instantaneous phase using the concept of analytic signals and by utilizing the Hilbert transform \citep{rosenblum1996phase}. For a given signal $x(t)$ normalized by its global maxima, the instantaneous amplitude $A(t)$ and phase $\phi(t)$ can be obtained from the complex analytic signal 
\begin{equation}
\xi(t)= x(t)+i\mathcal{H}\left(x(t)\right) = A(t)\exp\left(i\phi(t)\right)    \label{Eq-Analytic_signal_def}
\end{equation}
where $A(t)$ and $\phi(t)$ denote the instantaneous amplitude and phase of the analytic signal. The Hilbert transform is $\mathcal{H}(x(t))=1/\pi \int^\infty_{-\infty} x(\tau)/(t-\tau)d\tau$, where the integral is evaluated at the Cauchy principle value. The analytic nature of temporal coefficients is verified for different states by plotting the phase space trajectory of $\xi(t)$ and adjudging its center of rotation. For periodic signals, there is a unique center of rotation, allowing us to uniquely specify the phase of the signal. As instantaneous phase is a monotonically increasing or decreasing quantity, it is first wrapped to the interval $[-\pi,\pi]$. Finally, synchronization among pairs of signals are evaluated by measuring the instantaneous phase difference: $\Delta \phi_{x_1,x_2} = \phi_{x_1}(t) - \phi_{x_2}(t)$. Signals are considered to be phase locked when the difference $\Delta\phi_{x_1,x_2}$ becomes bounded to a small interval $\epsilon$ ($\leq 2\pi$) around the mean phase $C$, to wit, $|\Delta\phi_{x_1,x_2}(t)-C|=|\phi_{x_1}-\phi_{x_2}-C|\leq \epsilon$ \citep{pikovsky2001universal}.

The temporal interactions of the identified structures are further visualised using the phase portraits or Lissajous plots for pair of signals. The exact relation between the signals are inferred based on the dominant frequencies ($f_1, f_2$) of the signals and the mean phase difference ($\Delta \phi_m$). This is done by constructing the pair of analytical signals based on superposition of the dominant frequencies,
\begin{equation}
\begin{split}
x_1 &= A_1 \sin(2 \pi f_1 t + \Delta \phi_m) + A_2 \sin(2 \pi f_2 t + \Delta \phi_m),\\
x_2 &= B_1 \sin(2 \pi f_1 t) + B_2 \sin(2 \pi f_2 t),
\end{split}
\label{eq: 3.11}  
\end{equation}
where $f_1$ is the dominating frequency of the signal and $f_2$ is the multiple of $f_1$, i.e., $f_2$ = 2$f_1$. $A_1, A_2, B_1$ and $B_2$ are the amplitudes associated with the dominant modes $f_1$ and $f_2$ for different pair of signals. 

\subsection{Quantifying chaotic evolution of coherent structures}
\label{chaos test}

In order to quantify chaotic time evolution of coherent structures, we perform $0-1$ test which quantifies the unbounded growth of phase space trajectories typical of chaos \citep{gottwald2004new}. For any input time series $\gamma(t)$, the first step is to compute the translation variables $x(n)$ and $y(n)$ such that
\begin{equation}
   x(n) = \sum_{j=1}^n \gamma(t)\cos(tc), \enspace 
   y(n) = \sum_{j=1}^n \gamma(t)\sin(tc)
\end{equation}
where, $n = 1,2,...,N$, where $N$ denotes the number of data points present in time signal. For our analysis, we have selected the value of $c$ in the interval $(\pi/5, 4\pi/5)$ \citep{nair2013loss}. The nature of these two translation variables provide indications to determine if chaos exists in the system. For periodic or quasi-periodic oscillations, the variables show bounded behavior. However, for chaotic dynamics, their behavior is unbounded and irregular. The evolution of the trajectory in the $x-y$ plane for increasing $n$ can be computed with the help of modified mean square displacement $M(n)$ as follows \citep{ashwin2001hypermeander}:

\begin{equation}
\begin{aligned}
  M(n) = \frac{1}{N} \sum_{j=1}^n \left[(x(j+n)-x(j))^2+ (y(j+n)-y(j))^2\right] -\frac{1}{N} \sum_{j=1}^n\gamma(t)  \frac{1-\cos(jc)}{1-\cos(c)}
\end{aligned}
\end{equation}

For a chaotic state, the displacement $M(n)$ between the translational variables grows monotonically with $n$, while it becomes nearly constant for regular states. With the use of linear regression, the asymptotic growth rate ($K$) of the mean displacement is computed as follows:
\begin{equation}
\centering
K = \lim_{n\to \infty} \frac{\log M(n)}{\log n}.
\end{equation}

\cite{gottwald2009implementation} pointed that for the small values of \textit{n}, the linear regression methods results in altering the $K$ values. To estimate the $K$ value, we use a correlation method which performs better than the linear regression method. In this method, $K$ is the correlation coefficients for the vectors $\zeta = (1,2,3,...,n)^T$ and $\Delta = (M(1),M(2),M(3),...,M(n))^T$. Subsequently, the correlation coefficient can be defined as
\begin{equation}
    K = \textrm{corr}(\zeta,\Delta) = \frac{\textrm{cov}(\zeta,\Delta)}{\sqrt{\textrm{var}(\zeta)\textrm{var}(\Delta)}} \in [-1,1]
\end{equation}
where,
\begin{equation*}
    \textrm{cov}(\zeta,\Delta) = \frac{1}{n} \sum_{j = 1}^n \left[\zeta(j) - \Bar{\zeta}\right]\left[\Delta(j) - \Bar{\Delta}\right], \enspace \textrm{var}(\zeta) = \textrm{cov}(\zeta,\zeta), \enspace \textrm{and} \enspace \Bar{\zeta} = \frac{1}{n} \sum_{j=1}^n \zeta(j).
\end{equation*}
The value of $K$ lies between 0 and 1. For a chaotic signal, $K$ takes a value close to 1, and for a regular signal, it approaches a value close to 0. 

\section{Results and discussion}
\label{section:Results and discussion}

We begin by examining three distinct dynamical states - chaotic, period-1 and period-2 oscillations exhibited by the combustor at different levels of $\textrm{H}_2$ enrichment. The operating conditions corresponding to these states are highlighted in table \ref{table1}. The combustor is initially in a thermoacoustically stable state, operating only on $\textrm{CH}_4$ at a premixed equivalence ratio of $\phi= 0.65$ and a thermal power of 25.31 kW. During these conditions, thermoacoustic system is characterized with low-amplitude, chaotic pressure oscillations. We refer to this baseline state as chaotic state. When $\textrm{CH}_4$-air flame is enriched with $\textrm{H}_2$, the pressure fluctuations inside the combustor exhibits high amplitude period-1 and period-2 limit cycle oscillations (LCO) at 50\% and 20\% hydrogen enrichment. We refer to these states as period-1 and period-2 states, respectively. 

We disentangle large-scale coherent structures in the flow velocity field and their impact on the flame surface and heat release rate field by performing the POD and extended POD analysis. Large-scale structures in turbulent flows contain majority of the turbulent kinetic energy which commences the cascade process and turbulence phenomenology. Thus, POD modes with the largest kinetic energy content along with correlated flame structures would dominate the flow and flame dynamics. This can be observed simply by noting that the turbulent kinetic energy $E$ and fluctuating kinetic energy $E_i$ contained in each POD mode are connected via the relations between the eigenvalues of the correlation matrix \eqref{Eq-CorrelationMatrix} and the flow field fluctuations:

\begin{align}
E_i  \coloneqq \frac{1}{2}\overline{(\textbf{u}^\prime,\boldsymbol{\Psi}_i)^2}=\frac{1}{2}\lambda_i, \qquad E=\sum_{i=1}^N E_i=\frac{1}{2}\sum_{i=1}^N \lambda_i.
\label{Eq-TKE_defintion}
\end{align} 

\begin{figure}
%\setstretch{1.5}
\centerline{\includegraphics[width = 0.7\textwidth]{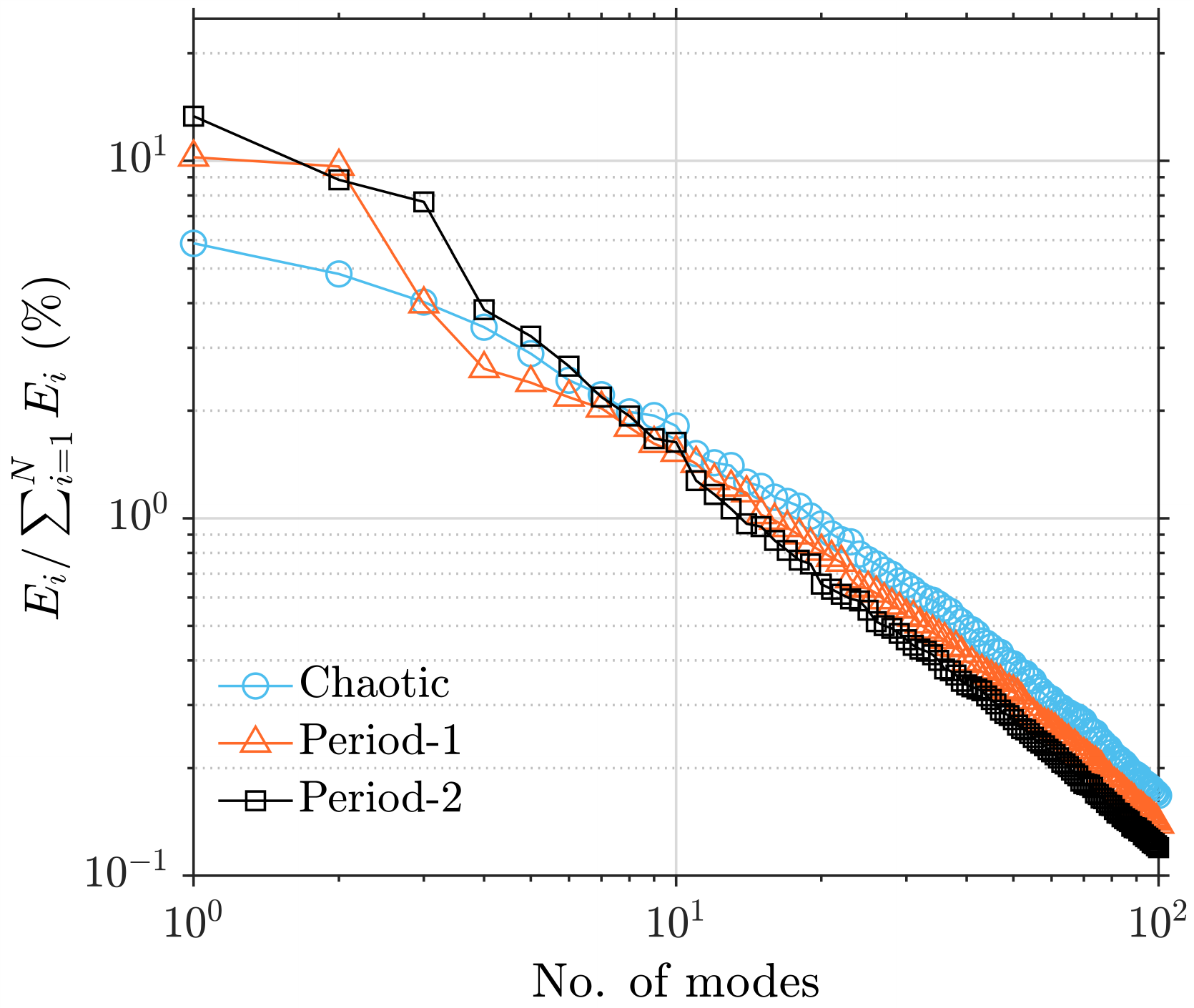}}% Images in 100% size
\caption{Fraction of kinetic energy contained in the spectrum of POD modes \eqref{Eq-TKE_defintion} obtained from the flow velocity field ($\textbf{u}^\prime$) for the three dynamical states highlighted in Table \ref{table1}. The key role played by large-scale coherent structures during the states of thermoacoustic instabilities (period-1 and period-2) is highlighted by the first two principal modes containing an order of magnitude higher kinetic energy than observed during stable operation.} 
\label{fig2}
\end{figure}

Figure \ref{fig2} shows the percentage of the turbulent kinetic energy ($E_i/E$) that the first 100 modes of the velocity field contain for different dynamical states. For all the cases, the first 100 modes contain over 90\% of the turbulent kinetic energy, with the first few principal modes accounting for a major fraction of the total kinetic energy. During chaotic oscillations, the first three POD modes contain $14.74\%$ of the total kinetic energy. Further, for period-1 LCO, the first two modes contain $19.87\%$ of the turbulent kinetic energy. The third mode contains an order of magnitude less fraction of the total energy. The energy content decreases drastically for higher POD modes. For period-2 LCO, the first three modes contain comparable proportion of the total kinetic energy, with the total amounting to around $29.86\%$ of the total energy.

In keeping with the definition, the spatial structures associated with the first few principal modes are expected to dominate flow field dynamics. For subsequent analysis and visualization, we consider the transverse component of the velocity field ($u_x$) and its effect on the turbulent flame. Our conclusions remain the same when the axial component ($u_y$) is considered instead (see supplementary information).

\subsection{Chaotic oscillations in the absence of H$_2$}
\label{subsection:CO1}

\begin{figure}
%\setstretch{1.5}
  \centering
  \includegraphics[width=0.75\textwidth]{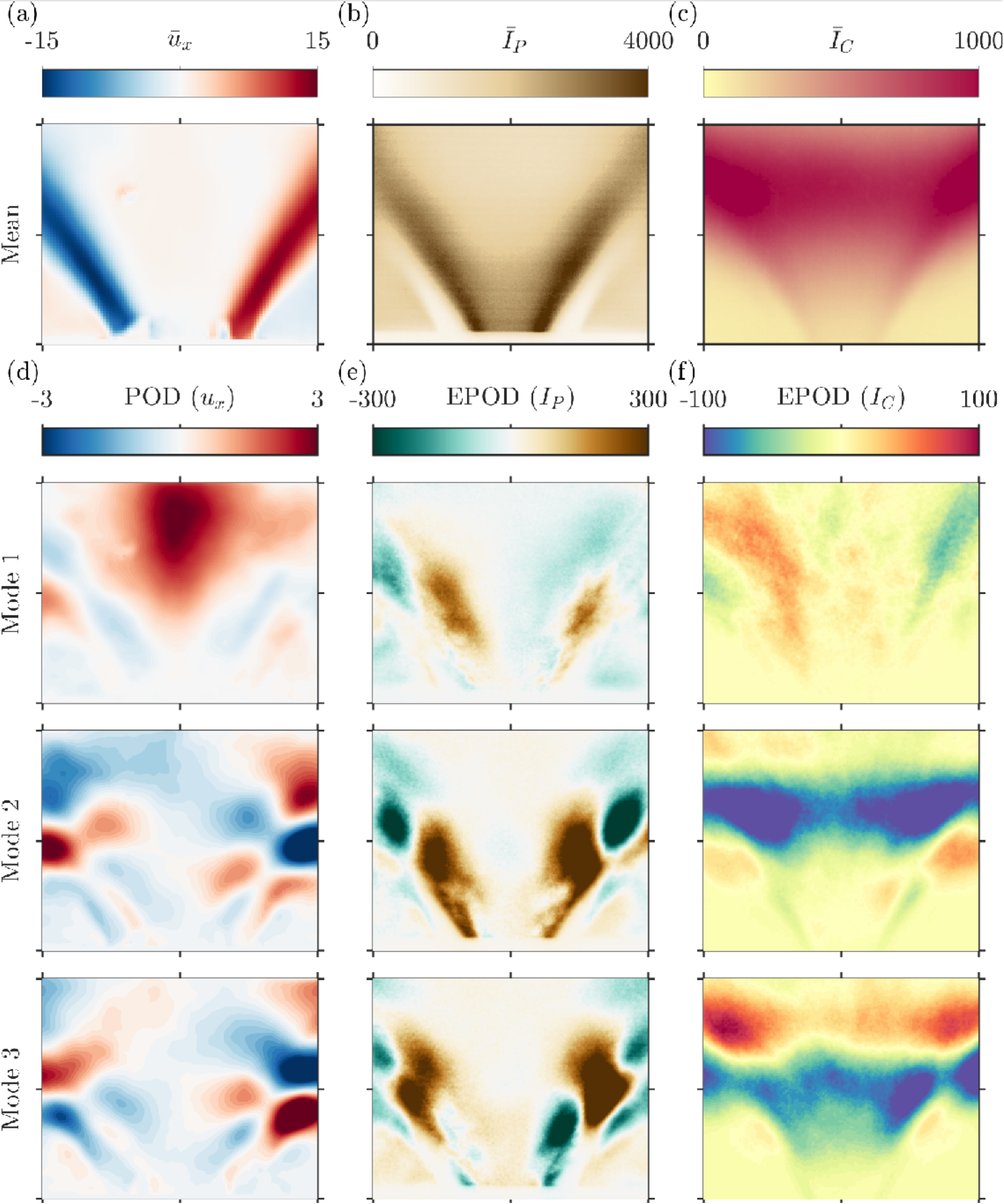}% Images in 100% size
  \caption{Flow and flame dynamics for the purely $\textrm{CH}_4$-air flame for the baseline condition. Mean of the (a) transverse velocity component $u_x$, (b) flame brush obtained from OH-PLIF images, and (c) heat release rate fluctuations obtained from OH*-chemiluminescence. Panels (d-f) shows the first three POD modes of $u_x^\prime$ and the extended modes associated with $I_P^\prime$ and $I_C^\prime$.}
\label{fig3}
\end{figure}

The flame and flow dynamics of the swirl combustor operating only on $\textrm{CH}_4$ at $\phi=0.65$ are shown in figure \ref{fig3}. The first row shows the mean of the transverse velocity component ($\bar{u}_x$), the flame brush obtained from time-averaged OH-PLIF imaging ($\bar{I}_P$) and the mean heat release rate obtained from OH*-chemiluminescence imaging ($\bar{I}_C$). The time averaging is performed over 3000 instantaneous images (or 0.3 seconds). Figure \ref{fig3}a shows high transverse velocity along the shear layers, with the opposite signs indicating the clockwise direction of the swirling flow. In addition, the flow shows very weak inner and outer recirculation (also seen from figure 2a showing $\Bar{u}_y$  in supplementary materials). Hence, the flame can be seen to stabilize along the inner shear layer, resulting in a characteristic V-shape of the turbulent flame brush (figure \ref{fig3}b). The thickness of the flame brush can also be seen to increase downstream in comparison to the attachment point at the exit of the dump plane. As chemiluminescence images are line-of-sight integrated in the out-of-plane direction, increased flame brush thickness can be seen to result in sustained, continuous band of heat release rate profile downstream of the combustor (figure \ref{fig3}c).

\begin{figure}
%\setstretch{1.5}
  \centering
  \includegraphics[width=\textwidth]{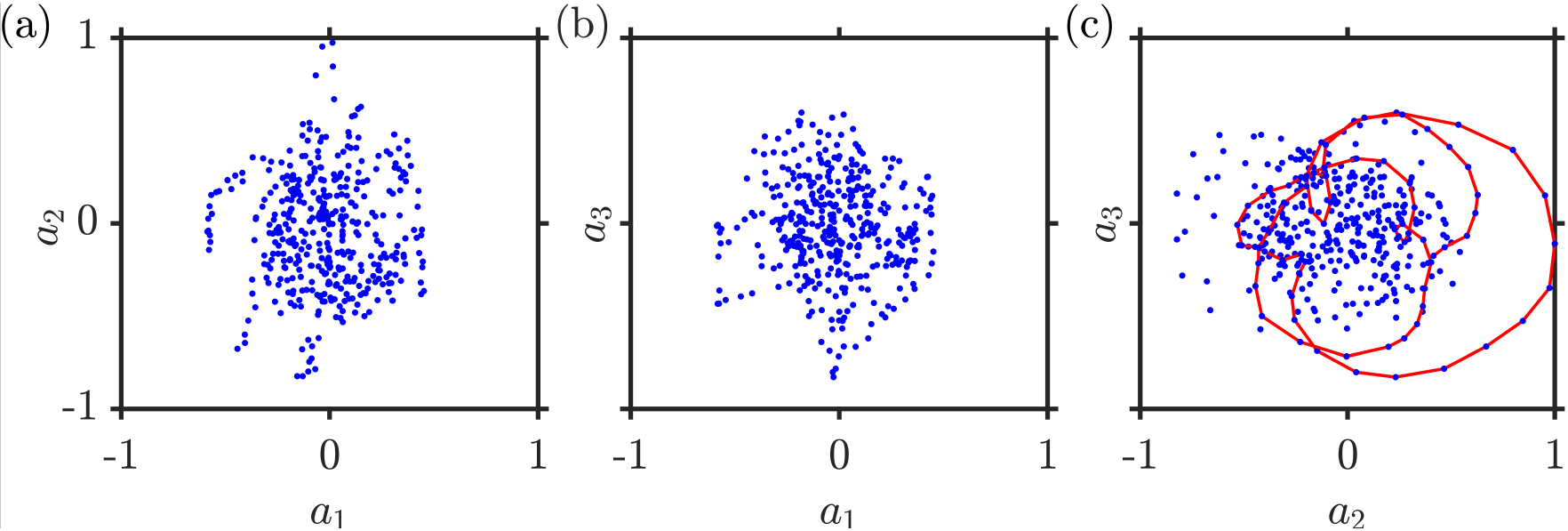}% Images in 100% size
  \caption{(a-c) Phase portrait of the pairs of temporal coefficients ($a_1$ - $a_3$) of the velocity field for the state of chaotic oscillations. The red lines indicate a few periodic orbits.}
\label{fig41}
\end{figure}

Figure \ref{fig3}d shows the first three POD modes of the transverse ($u_x$) velocity, together which account for $14.74\%$ of TKE energy (figure \ref{fig2}). The first POD mode of the transverse velocity shows a coherent structure that grows in magnitude in the streamwise direction. The structure indicates a vortex bubble with positive velocity magnitude, depicting the overall direction of its motion. In contrast, mode 2 and mode 3 depict modes with same wavenumber and frequency, whose magnitude increases and then decays downstream. These structures are shifted in the streamwise direction by approximately quarter wavelength. Mode 2 and 3 are part of an oscillating process, linked to a helical mode instability around the recirculation bubble observed in POD mode 1. This can be observed in the Lissajous plot between mode coefficients $a_2-a_3$ (figure \ref{fig41}). We further note here that the first three POD modes associated with $u_y$ along the shear layer (in figure 2d in supplemental materials) also indicate the presence of the same coherent structures with the same axial wavenumber with each mode shifted axially. Additionally, all the three modes exhibit advection of the central recirculation zone (in figure 2d in supplemental materials). 

The extended POD modes of the flame profile ($I_p$) associated with the POD modes of $u_x$ are also shown in figure\ref{fig3}e. We notice a number of salient features of the flame dynamics from the extended POD modes of the flame surface $I_p$ in figure \ref{fig3}e. The dominant POD mode of $u_x$ associated with the central recirculation bubble, does not correlate with extended POD modes of the flame surface. Only the weaker structures along the shear layer are reflected in the extended POD modes of the flame surface. This is due to the V-shaped flame (figure \ref{fig3}b) which stabilizes along the inner shear layer and does not extend into the inner recirculation zone. The increased activity in the flame surface is further emphasized in the second and third extended POD modes of the flame surface. The coherent structures present along the shear layer in the second and third POD modes of $u_y$ are clearly correlated to the extended POD mode structures. The extended POD mode structure is also shifted by a quarter wavelength. The amplitude of the structure can be see to increase downstream of the flame attachment region, identifying regions contributing to the most of the changes in the flame dynamic. Interestingly, while the coherent structure is anti-symmetric (mode 2 \& 3 in figure \ref{fig3}d) due to helicoidal instability in the flow field, the extended mode structure of the flame surface clearly retains axisymmetry. 

\begin{figure}
%\setstretch{1.5}
  \centering
  \includegraphics[width=\textwidth]{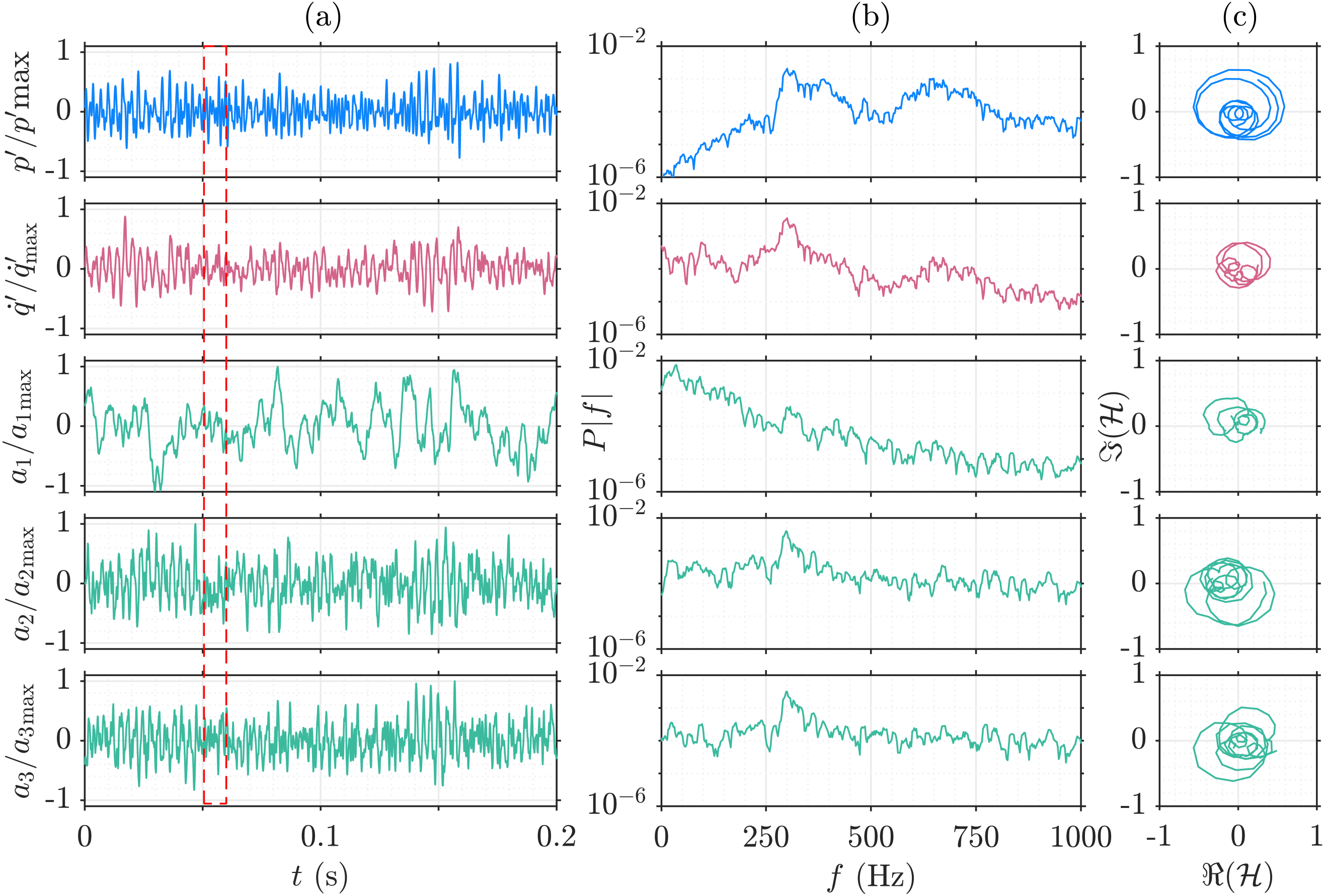}% Images in 100% size
  \caption{Characterization of the thermoacoustic response during the state of chaotic oscillations. (a) Time series and (b) power spectra of fluctuations in acoustic pressure, heat release rate and the temporal coefficients of the first three POD modes. (c) Phase space trajectory in the complex analytical plane for the part of the signals indicated in panel (a). }
\label{fig4}
\end{figure}

The effect of these local flow and flame dynamics on the global heat release rate profile can be understood from extended mode structure of heat release rate distribution shown in figure \ref{fig3}f. As with the extended structure of flame surface, the first extended mode of heat release rate does not correlate with the central recirculating zone, corroborating the fact that the central vortex bubble arising due to the swirl does not affect the fluctuations in the heat release rate. The second and third extended modes on the other hand correlate well with the coherent structure present in the velocity field. Indeed, we notice that structure of extended POD of heat release exists along the shear layers along with the extended mode structure of the flame surface. The second and third mode are again related to the same structure, only displaced by a quarter wavelength. The banded nature of these structures can be ascribed to the fact that the chemiluminescence imaging is line-of-sight integrated, revealing the annular region over which heat is release from the three dimensional flame structure. We note here that our observations remain the same when the streamwise velocity $u_y$ is used for obtaining extended POD modes from flame surface and heat release rate (see figure 2e,f in supplemental materials). These observations show that the structures from the heat release rate distribution and flame surface corresponding to the most dominant POD mode do not correlate. The coherent structures from other modes show some correlation with the structures from other fields. However, due to low turbulent kinetic energy of mode 2 and 3, their overall effect remains minimal. 

\begin{table}
\begin{center}
%\def~{\hphantom{0}}
%\makegapedcells
\centering
    \begin{tabular}{|c|c|c|c|}
       \backslashbox{Signals $\downarrow$}{Dynamical states $\rightarrow$}  &  Chaotic oscillations & Period-1 LCO & Period-2 LCO \\
      \hline
       $p^{\prime}$ & 0.99 & 0.08 & 0.01 \\
       $\dot{q}^{\prime}$   &   0.98 & 0.11 & 0.05 \\
        $a_1$ &   0.62 & 0.15 & 0.27 \\
       $a_2$ &   0.99 & 0.12 & 0.32 \\
       $a_3$ &   0.99 & 0.61 & 0.79 \\
    \end{tabular}
  \caption{Results of the 0-1 test for different signal during all the dynamical states discussed in this paper. The values close to 1 indicate the existence of chaos in the signal. Moreover, the signal with regular dynamics show the values close to 0. }
  \label{table2}
  \end{center}
\end{table}

We next relate the manner in which these spatial modes are related to the thermoacoustic response in the system. Figure \ref{fig4} presents the normalized time series of the  acoustic pressure ($p^\prime$) and heat release rate fluctuations ($\dot{q}^\prime$) along with the time evolution of the temporal coefficients of the first three POD modes. Here, $\dot{q}^\prime(t)$ is obtained by a global summation from individual chemiluminescence images. Each of the time series are normalized by their respective maxima to aid comparison. The aim is to relate the manner in which the spatial modes evolve in time and relate to fluctuations in $p^\prime$ and $\dot{q}^\prime$. In each case, the time evolution is characterized by the power spectral density and the phase portrait of the analytic signal defined according to \eqref{Eq-Analytic_signal_def}.

The present baseline case corresponds to the low-amplitude stable operation of the combustor. The pressure and heat release rate oscillations during this baseline case remains chaotic, as reported earlier in \citep{kushwaha2021dynamical}. This can be observed in figure \ref{fig4}b where the power spectral density for $p^\prime$, $\dot{q}^\prime$ and the temporal coefficients ($a_1, a_2, a_3$) depict a broadband behavior. In addition, the phase portrait featuring the analytic representation of these signals depict the absence of a unique center of rotation. Hence, the signals are non-analytic, and a phase cannot be ascribed to these signals.

We perform the 0-1 test (described in \S\ref{chaos test}) for these signals to ascertain whether their dynamics is chaotic or not. The value of the asymptotic growth rate $K$ of the displacement $M$ between translational variables is tabulated in Table \ref{table2}. We note here that all the signals display values which are close to unity, implying chaotic evolution of each of these quantities. Thus, the heat release rate and pressure oscillations display chaos during stable combustor operation, corroborating past results \citep{nair2014multifractality, kushwaha2021dynamical}.  

\subsection{Period-1 limit cycle oscillations at 50\% $\textrm{H}_2$ enrichment}
\label{subsection:P1-LCO}

\begin{figure}
%\setstretch{1.5}
  \centerline{\includegraphics[scale = 0.56]{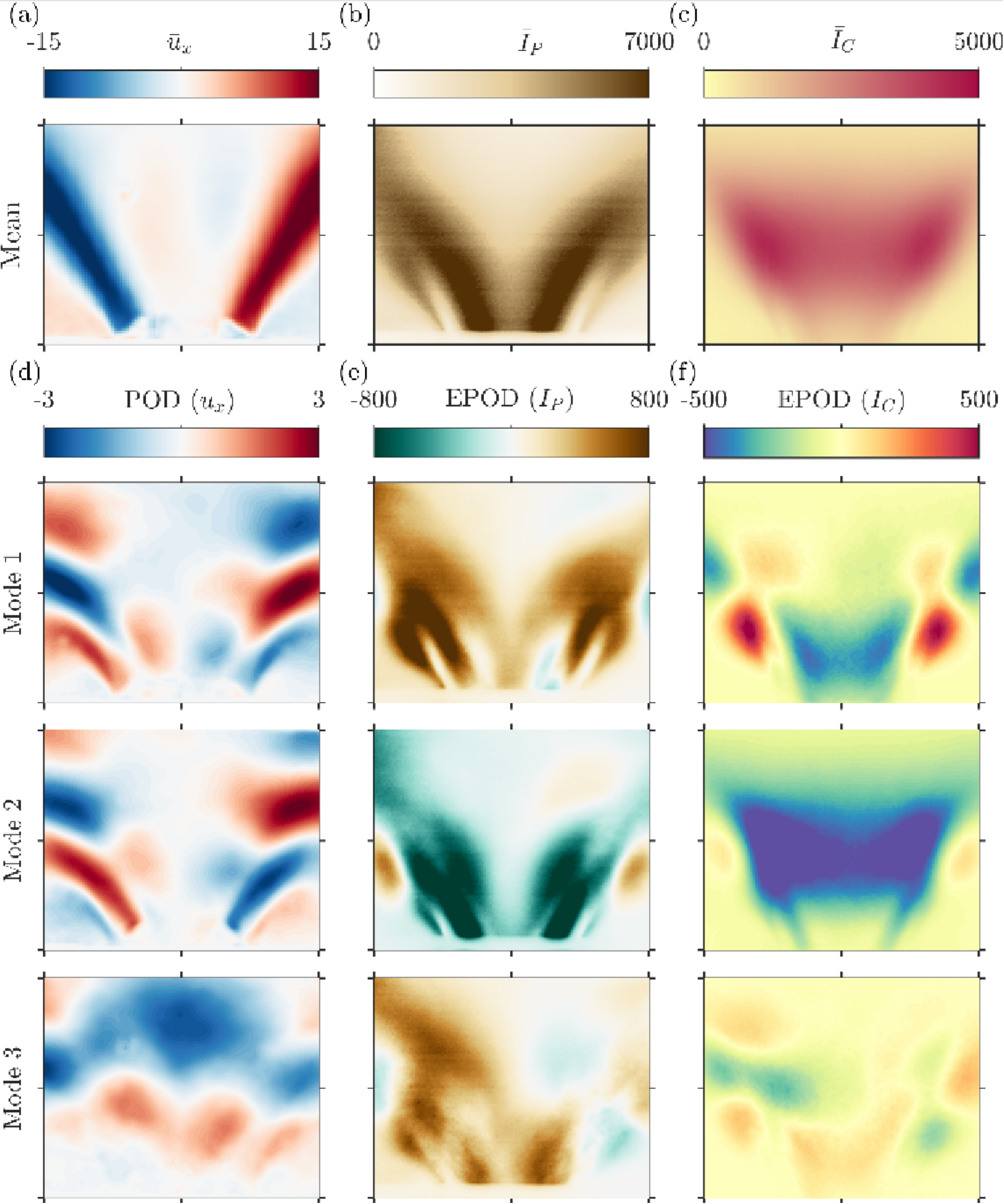}}% Images in 100% size
  \caption{Flow and flame dynamics during Period-1 TAI with 50\% hydrogen enrichment. Mean of the (a) transverse velocity component $\bar{u}_x$, (b) flame brush obtained from OH-PLIF images, and (c) heat release rate fluctuations obtained from OH*-chemiluminescence. Panels (d-f) shows the first three POD modes of $u_x^\prime$ and the extended modes associated with $I_P^\prime$ and $I_C^\prime$.}
\label{fig5}
\end{figure}

The combustor shows the state of TAI with the addition of hydrogen (50\%) at a constant equivalence ratio of $\phi=0.65$ and a thermal power of $P=20$ kW. We observe sustained, period-1 LCO with an amplitude of $p^\prime_{\textrm{rms}}=0.91$ kPa. \Figref{fig5} shows the time-averaged velocity field, flame brush and heat released rate. \Figref{fig5}a shows high transverse velocity along the shear layers having opposite sign which indicates swirling flow. The shear layer is much broader in comparison to the baseline case. The flame also stabilizes along the shear layers and extends towards the outer recirculation zone, making it an M-shaped flame (figure \ref{fig5}b). The flame can be seen to be anchored very close to the dump plane. The time-averaged image of the heat release distribution field shows a concentrated region of high heat release rate. The location of the peak of $\bar{I}_C$ is located much closer to the flame anchor point, in contrast to the baseline chaotic case where the peak occurred much further downstream.

Figure \ref{fig5}d shows the first three dominant POD modes associated with the transverse velocity fluctuations $u_x^\prime$. These three modes collectively hold $30\%$ of the total kinetic energy. The first and second POD modes indicate the presence of the same coherent structure in the form of a toroidal vortex of the same frequency and wavenumber. The modes can be seen to be shifted by a quarter of a wavelength and $\pi/2$ radians (see figure \ref{fig8}a). In contrast to the baseline case, the dominant coherent structure during period-1 LCO comprises the helical toroidal vortex. The amplitude of the toroidal vortex can be observed to be much stronger than it was for the baseline case (cf. figure \ref{fig3}). The toroidal vortex can also be seen to extend to a much larger radial domain, with the anti-phase features alternating asymmetrically on each side of centre flow axis ($x = 0$). In addition to the toroidal vortex, the first two POD modes also show the presence of an anti-symmetric mode structure along the IRZ which convects in time. The third mode shows the presence of an axisymmetric coherent structure extending along the length of the combustor. 

\begin{figure}
%\setstretch{1.5}
  \centering
  \includegraphics[width=\textwidth]{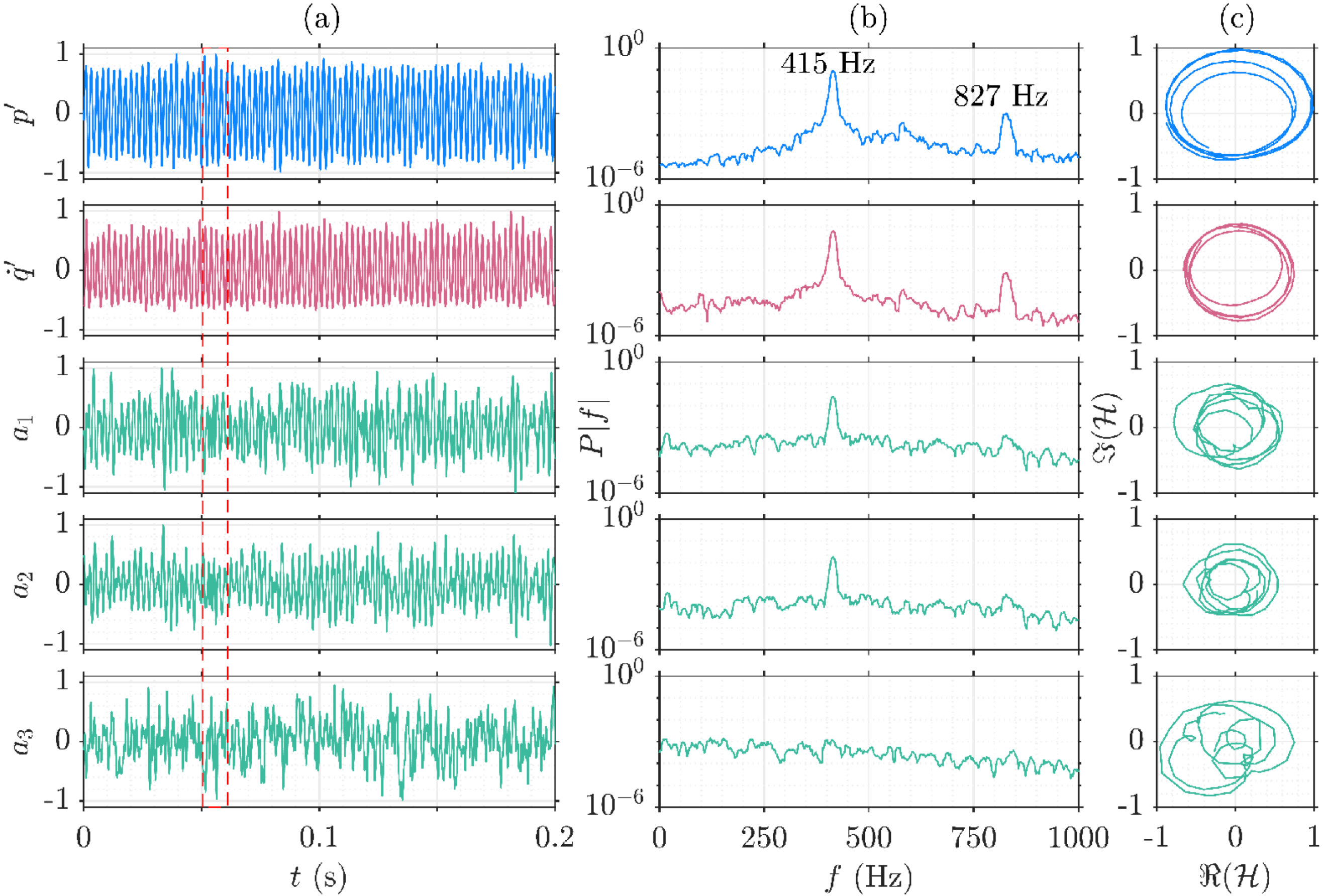}% Images in 100% size
  \caption{Characterization of the dynamic response during period-1 TAI. (a) Time series and (b) power spectra of acoustic pressure, heat release rate and temporal coefficients of first three POD modes. The time series are normalised by their respective maxima values. (c) Phase space trajectory in the complex analytical plane for the part of the signal indicated in panel (a). }
\label{fig6}
\end{figure}

\begin{figure}
%\setstretch{1.5}
  \centerline{\includegraphics[width=\textwidth]{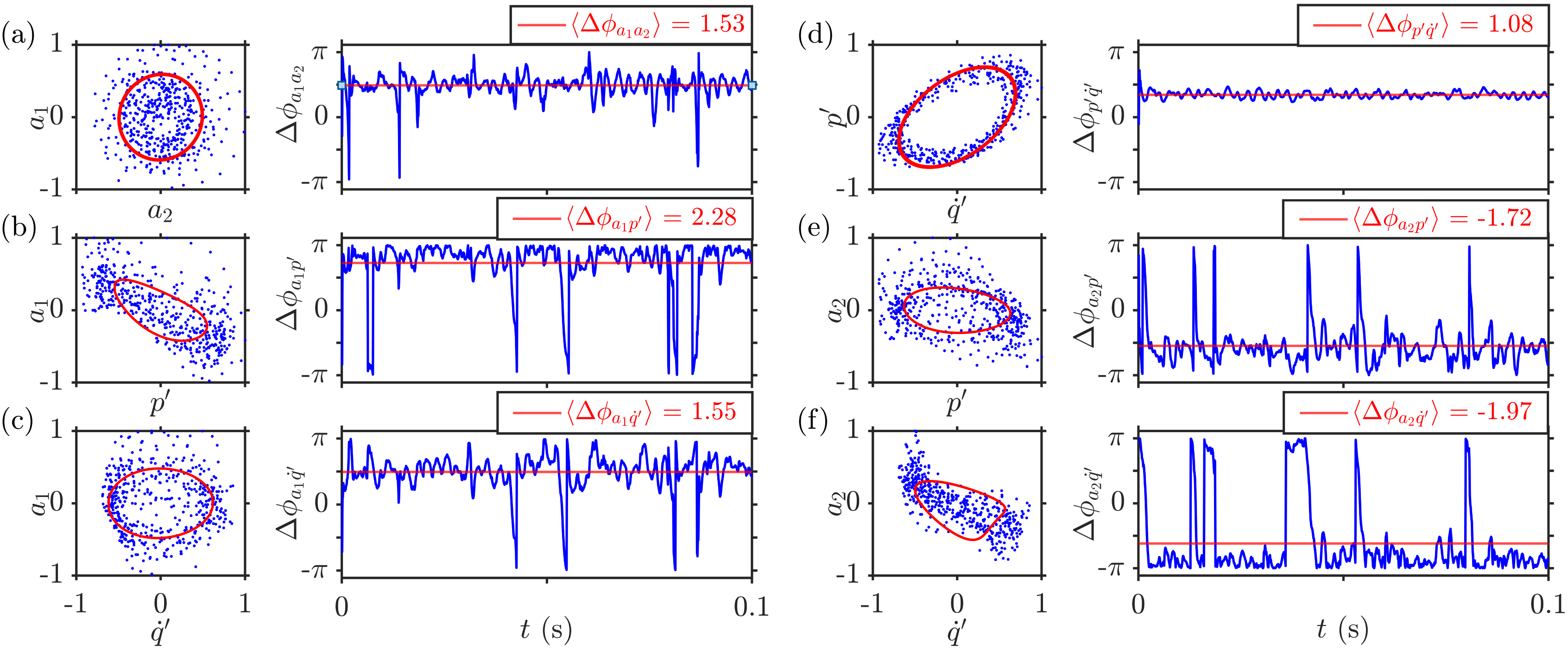}}% Images in 100% size
  \caption{(a-f) Phase portrait of the temporal coefficients of the flow velocity field with respect to acoustic pressure and heat release rate along with analytical curves having the same dominating frequencies and phase differences as the original data and the corresponding phase difference with the red line indicating the mean value of phase difference, for the state of TAI with period-1 LCO. }
\label{fig7}
\end{figure}

The extended POD modes of the flame surface is shown in figure \ref{fig5}e. For the first two modes, we notice that the extended structure is distinctly M-shaped with branches of the flame surface extending along the nodal line of the toroidal as well as the inner coherent structure. The two extended modes, which are shifted by $\pi/2$ radians, are associated with positive and negative intensities, indicating the correlation between POD mode and the extended POD mode structure of the flame surface. In addition, the first and the second EPOD modes have the maximum intensity along the shear layers. However, we observe small scale structures in the third EPOD mode structure of the flame. Figure \ref{fig5}f present the EPOD modes of the heat release rate distribution. The first two EPOD modes show the structures in the IRZ and along the shear layer. The structures along shear layer confirms the existence of a toroidal vortex in the flow field (figure \ref{fig7}d). However, we do not observe the high intensity structure in the third EPOD modes of the OH-PLIF images and OH*-chemiluminescence. With the random arrangement of structures in the third extended modes of the flame surface and heat release rate, the coherent structures from the flow field do not correlate with the inner recirculating zone. 

To probe the dynamics further, the time series, power spectra and analytical signal are shown in figure \ref{fig6}. The time series of $p^{\prime}$ and $\dot{q}^{\prime}$ signals, normalised with their respective global maxima values, show high amplitude fluctuations compared to that during the chaotic oscillations (figure \ref{fig6}a). We also observe a distinct peak at 415 Hz and its higher harmonic at 827 Hz for the $p^{\prime}$ and $\dot{q}^{\prime}$ signals (figure \ref{fig6}b). The temporal coefficients $a_1$ and $a_2$ are periodic in nature (figure \ref{fig6}a) and exhibit distinct peak at 415 Hz (figure \ref{fig6}b). However, the third mode does not show periodic behaviour (figure \ref{fig6}a) and we observe a broadband power spectrum without any distinct peak (figure \ref{fig6}b). However, the temporal coefficient of the third mode shows aperiodic behaviour (figure \ref{fig6}a) and exhibit a broadband power spectrum without any distinct frequency peak (figure \ref{fig6}b). The trajectory on the complex plane also does not have a unique center of rotation, indicating that the signal is not analytic. Further, from Table \ref{table2}, we notice that the $0-1$ test of $a_1$ indicates a value of $K=0.61$, indicating a possible chaotic evolution of mode 3.

In order to understand the coupled behaviour between the variables, we show the Lissajous plots and determine the phase difference between various pairs of signals in figure \ref{fig7}. The phase space trajectory plotted in the complex analytic plane, shown in figure \ref{fig6}c, clearly depicts a unique center of rotation for all the pair of signals except for $a_3$. This indicates that all signals other than $a_3$ are analytic in nature and their phases are well-defined. We compute the mean of the instantaneous phase difference and show it with a red line. With the help of dominant frequencies and the mean phase difference for each pair of signal, we generate the analytical curve according to \eqref{eq: 3.11} and depict it in the Lissajous plot. These analytical curves identify the phase and frequency relationship among the dominant POD modes and the pressure and heat release rate fluctuations. We further note that the pressure fluctuations correspond to the acoustic standing wave of the combustor, ensuring that the phase jumps by 2$\pi$ when the nodal line is crossed. Thus, the phase relationships discussed next are expected to remain unchanged close to the dump plane of the combustor where the POD are shown in figure \ref{fig6}.

In figure \ref{fig7}a, the phase plot between $a_1$ and $a_2$ shows that the first two POD modes are phase-locked around $\pi/2$ and are part of the same oscillating coherent structure. We also notice that $p^\prime$ and $\dot{q}^\prime$ are phase-locked close to a mean value of $\langle\Delta \phi\rangle = \pi/3$ (figure \ref{fig7}d), implying that the Rayleigh criteria is fulfilled during the state of TAI. $\langle\cdot\rangle$ indicates a time-averaged value. The relationship between the dominant POD modes and $p^\prime$ and $\dot{q}^\prime$ are quite interesting. We note that $a_1$ is phase-locked to $p^\prime$ at $\Delta \phi= 3\pi/4$ denoted by the elliptical Lissajous plot (figure \ref{fig7}b) and to $\dot{q}^\prime$ at $\Delta\phi=\pi/2$ as denoted by the circular plot (figure \ref{fig7}c). Next, we notice that the second mode $a_2$, which is associated with the same dominant coherent structure, is phase-locked to $p^\prime$ at $-\pi/2$ (figure \ref{fig7}e) and with $\dot{q}^\prime$ with a phase difference of $\approx -5\pi/8$.

One of the mechanisms through which the state of TAI occurs is often related to the presence of coherent structures. Previous researches have shown that the heat release rate fluctuations induced by the coherent structures are in-phase with the pressure fluctuations \citep{poinsot1987vortex, chakravarthy2007vortex, pawar2017thermoacoustic}. However, the phase relationships presented above suggest a non-trivial manner in which the coherent modes interact with $p^\prime$ and $\dot{q}^\prime$ fluctuations. The present relationship suggests that most energetic coherent structures evinces heat release rate fluctuations delayed by a phase-lag of $\pi/2$, which are subsequently related to $p^\prime$ fluctuations after a phase delay of $\pi/3$. However, the mismatch between the mean-phase relationship obtained from $\langle\Delta\phi_{a_1,p^\prime}\rangle$ and $\langle\Delta\phi_{a_1,\dot{q}^\prime}\rangle$ with $\langle\Delta \phi_{p^\prime,\dot{q}^\prime}\rangle$ indicates additional mechanism through which the coherent structure is coupled to the pressure fluctuations.

Thus, we have identified the phase-relationships between the coherent structures containing the most of the energy, flame fluctuations, global heat release rate fluctuations and the pressure field. The structures from the flame surface at the central plane and heat release rate distribution field are highly correlated with the coherent structures from the velocity field and temporally shows the 1:1 frequency locking behaviour among the subsystems in the combustor.  

%patterns (circles and ellipses) are similar to the Lissajous patterns which provides the frequency locking in terms of ratio between the dominating frequencies of the pair of the signals. We see differences between the patterns from our experimental data and the traditional Lissajous patterns due to existence of higher harmonics which affects the shape as well as the size of the patterns. These results identify 1:1 frequency locking behaviour among the acoustic pressure, the heat release rate and the flow field dynamics during the state of thermoacoustic instability with period-1 limit cycle oscillations. 

\subsection{Period-2 limit cycle oscillations (P2 LCO)}
\label{subsection:P2-LCO}

\begin{figure}
%\setstretch{1.5}
  \centerline{\includegraphics[scale = 0.58]{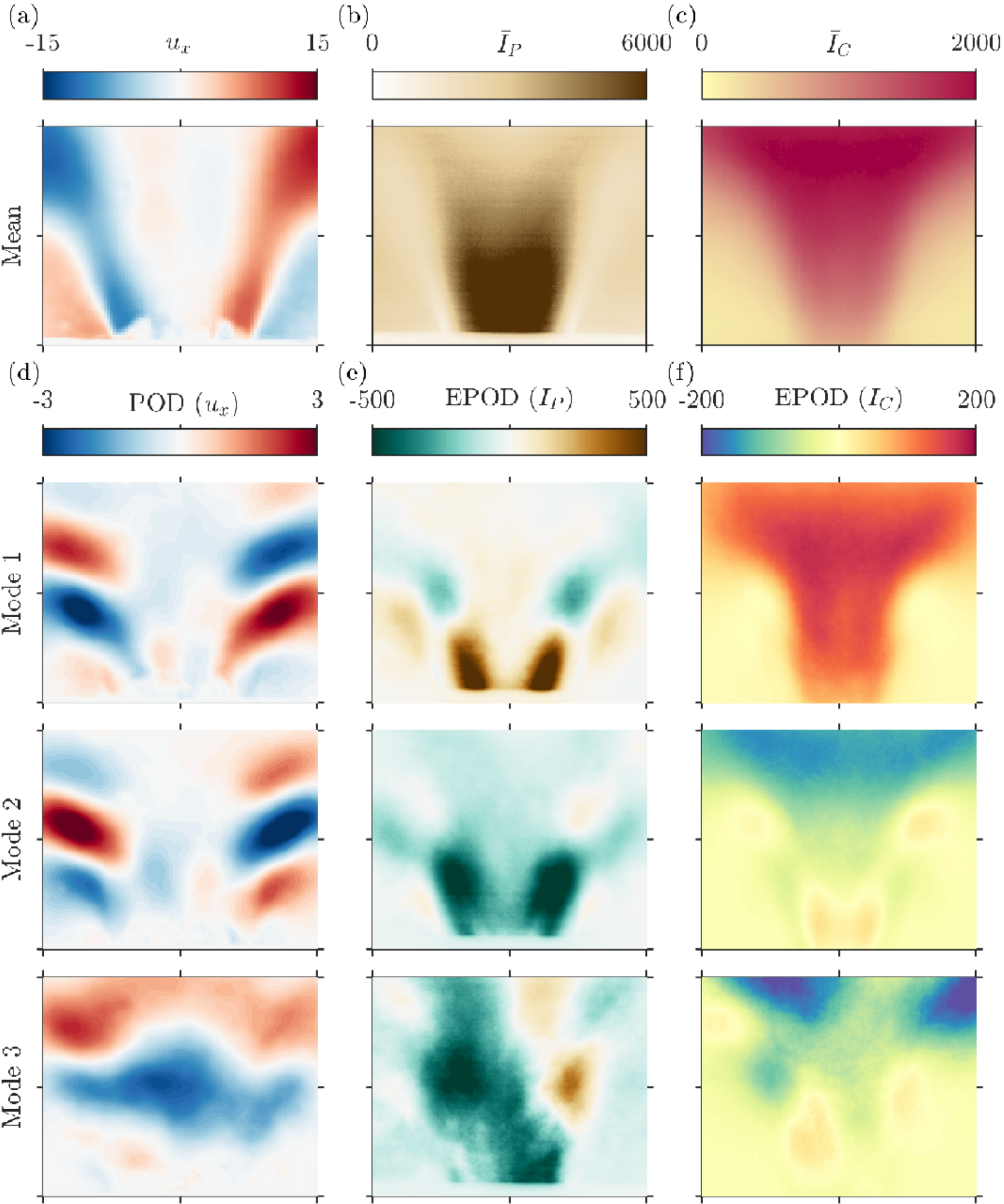}}% Images in 100% size
  \caption{Mean of the (a) transverse velocity component $u_x$, (b) flame brush obtained from OH-PLIF images, and (c) heat release rate fluctuations obtained from OH*-chemiluminescence. Panels (d-f) shows the first three POD modes of $u_x^\prime$ and the extended modes associated with $I_P^\prime$ and $I_C^\prime$. for the state of period-2 LCO with 20\% hydrogen by volume of the fuel. }
\label{fig8}
\end{figure}

\begin{figure}
%\setstretch{1.5}
  \centering
  \includegraphics[width=\textwidth]{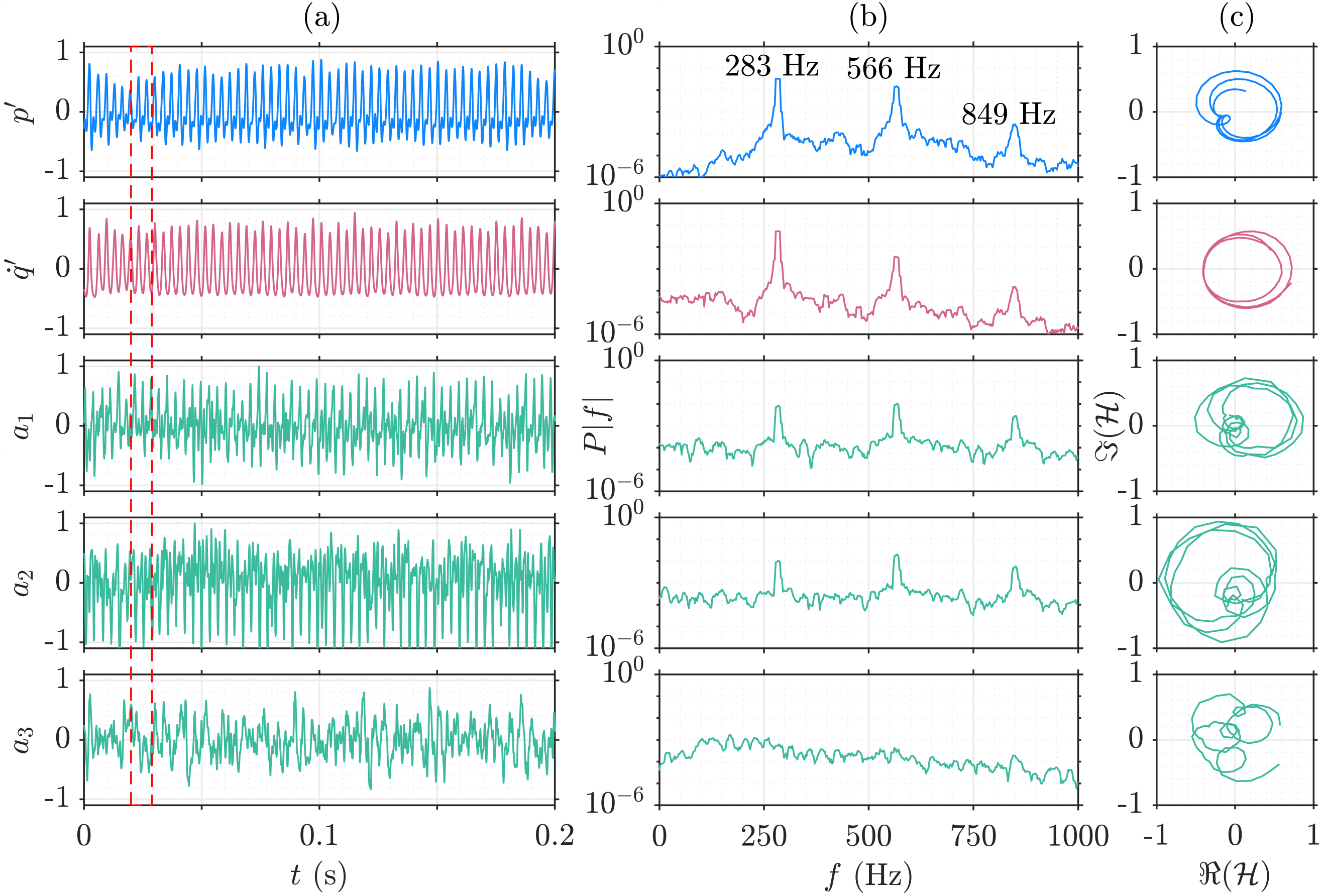}% Images in 100% size
  \caption{(a) time series of acoustic pressure, heat release rate and temporal coefficients of first three POD modes normalised by their respective maxima values; (b) the corresponding power spectra and (c) the analytical signals in the complex plane, corresponding to the time series enclosed in the red dashed box during the state of period-2 LCO.  }
\label{fig9}
\end{figure}

Let us now turn our attention to the coupled dynamics observed during the state of TAI with period-2 LCO when the combustor is operated at a constant equivalence ratio $\phi=0.65$ and thermal power $P=20$ kW with only $20\%$ hydrogen. Like earlier, we plot the mean, POD modes and extended POD modes in figure \ref{fig8} and the temporal dynamics in figure \ref{fig9}. 

In the mean transverse velocity field (figure \ref{fig8}a), we notice high velocity fluctuations along the shear layers and the outer recirculation region (ORZ) with the opposite signs. The strength of fluctuations are lower than that observed during period-1 LCO (figure \ref{fig5}a). Moreover, the mean of the OH-PLIF image shows the presence of a columnar shaped flame having very high intensity stabilized along the inner recirculation zone of the combustor (figure \ref{fig8}b). Most of the heat release rate occurs along the inner recirculation zone and extends downstream of the combustor as shown in figure \ref{fig8}c. This is in contrast to the mean profile observed for the M-flame during period-1 oscillations (figure \ref{fig5}a). These results indicate that the flame surface shape and the heat release rate distribution are entirely changed from the state of period-1 LCO with small variation in the volume of hydrogen to the fuel. 

To study the state of period-2 LCO thoroughly, the first three dominating POD modes of the transverse component are presented in figure \ref{fig8}d which cumulatively contains $23\%$ of the total kinetic energy. The first two modes show that the coherent structures are symmetrically positioned on each side of the flow axis and have anti-phase behaviour. These coherent structures confirm the existence of the helical nature of the flow. Moreover, very weak small-scale coherent structures with low velocity fluctuations can also be seen in IRZ. Unlike during period-1 LCO (figure \ref{fig5}d), the coherent structure can be seen to be limited only along the periphery and do extend inwards along the shear layer. However, the third mode shows the presence of a convecting structure along the axial direction (figure \ref{fig8}d). We also see similar elongated coherent structures in the POD modes of the axial velocity component (see figure 4 in supplementary material). These observations shows that the flow structures from the dominant modes confirm the existence of helical instability. However, the heat release rate field indicate that the maximum heat release occurs in the IRZ and does not follow the path of the helical structures in the flow. %which could be due to the longitudinal instability.

The EPOD modes of the flame surface corresponding to the first and second POD modes of velocity field exhibit structures with high flame surface fluctuations along the shear layer near to the inlet of the combustor (figure \ref{fig8}e). That indicates a longer flame surface than that of in case of period-1 LCO due to low volume of hydrogen as combustion with higher volume has smaller flame \citep{zhen2012effects}. The EPOD modes of the flame surface correlate well with the convecting coherent structure. Moreover, the third EPOD mode shows an asymmetrically distributed flame surface, convecting downstream with the flow (figure \ref{fig8}e). We also see similar structures near the IRZ in the extended modes of the flame surface using the axial velocity component of the flow (figure 4 in Supplementary material). The first two EPOD modes of heat release rate distribution confirm that the maximum heat release rate occurs downstream of the combustor (figure \ref{fig8}f). The first two EPOD modes of heat release rate distribution are symmetrical about the axis at $x = 0$. Similar to the third EPOD mode of the OH-PLIF images, the structures occur without any pattern in the third EPOD mode of heat release rate distribution field. The results from the first two modes clarify that the spatial locations of the coherent structures do not match with the locations of structures from EPOD modes. The coherent structures in the POD modes of axial velocity component shares the spatial positions with the structures from the corresponding EPOD of flame surface and heat release rate distribution fields (figure 4 in Supplementary material). Moreover, the structures from EPOD modes of flame surface and heat release rate distribution fields match.

%which indicate that different components of velocity affect the same structures in flame and heat release rate distribution fields. 

To understand the time dynamics of the state of period-2 LCO, we show the time series, power spectra and analytical signal in figure \ref{fig9}. The normalised time series of $p^{\prime}$ and$\dot{q}^{\prime}$ signals show high amplitude fluctuations. The acoustic pressure shows period-2 LCO, characterised by two dominant timescales. We observe that the signal repeats itself in every two oscillations \citep{kushwaha2021dynamical}. The heat release rate signal shows period-1 LCO (figure \ref{fig9}a). The power spectra of $p^{\prime}$ and $\dot{q}^{\prime}$ show a distinct peak at 283 Hz with its higher harmonics at 566 Hz and 849 Hz. The fundamental frequency and first harmonic in the $p^{\prime}$ power spectrum have the same order of power (figure \ref{fig9}b). On the complex plane, the trajectory of the analytical signal forms a double-looped pattern. We observe a unique center of rotation for the trajectory for  $\dot{q}^{\prime}$ (figure \ref{fig9}c) indicating that the signals are analytic. The temporal coefficients $a_1$ and $a_2$ have period-2 behaviour (figure \ref{fig9}a) and exhibit distinct peaks at 283 Hz  with higher harmonics at 566 Hz and 849 Hz (figure \ref{fig9}b). The signals $a_1$ and $a_2$, exhibiting the centers of rotation on the complex plane, are analytic (figure \ref{fig9}c). In case of the third mode, we do not see a periodic behaviour in the time series, as can be seen in figure \ref{fig9}a. Further, we notice a broadband spectrum without any distinct peak in the power spectrum (figure \ref{fig9}b). The trajectory of the signal does not have a unique center of rotation (figure \ref{fig9}c) which suggests that the signal is not analytic.    

\begin{figure}
%\setstretch{1.5}
  \centerline{\includegraphics[width=\textwidth]{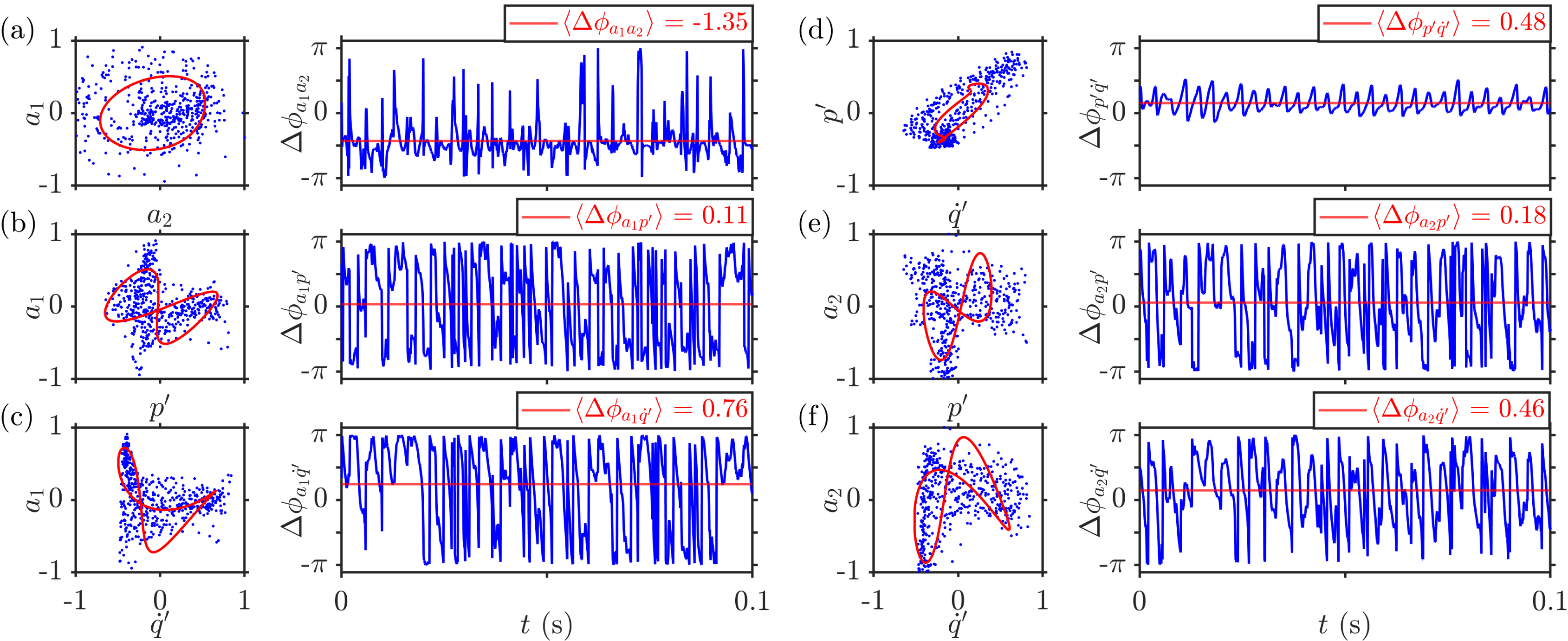}}% Images in 100% size
  \caption{(a-f) Phase portrait of the temporal coefficients of the flow velocity field with respect to acoustic pressure and heat release rate along with analytical curves having the same dominating frequencies and phase differences as the original data and (i - vi) the corresponding phase difference with the red line indicating the mean value of phase difference, for the state of TAI with P2 LCO.}
\label{fig10}
\end{figure}

In figure \ref{fig10}, we show the phase portraits for different pairs of signals and the phase difference between the signals for the corresponding pairs for the state of TAI with period-2 LCO. Unlike the state of period-1 LCO, we observe different patterns similar to the Lissajous figures. The phase portrait for the pair of the signals $a_1$ and $a_2$ shows that the coherent structures are phase-locked with the phase difference around $\pi/2$ (figure \ref{fig10}a). The reconstructed curve exhibit the circular pattern which confirms the 1:1 frequency locking between $a_1$ and $a_2$. The pair of the signals $a_1$ and $p^\prime$ show the phase locking behaviour with the mean phase difference between 0 and $\pi/4$ (figure \ref{fig10}a). However, the instantaneous phase difference between $a_1$ and $p^\prime$ varies a lot that could be due to the presence of high turbulence in the flow. The reconstructed curve in the phase portrait show a double-lobbed pattern (figure \ref{fig10}a) which is same as the Lissajous patterns indicating 2:1 frequency locking behaviour. Similarly, other pairs of signals also exhibit the patterns with the two lobes (figures \ref{fig10}c-f) which are traditionally observed for 2:1 frequency locking between the signals. The traditional 2:1 Lissajous patterns has lobes with equal shape and size (at least symmetric about one axis). However, we observe patterns with unequal shape and size due to the presence of higher harmonics in the experimental data. These results confirm that the temporal dynamics has both 1:1 and 2:1 frequency locking behaviour during the state of TAI with period-2 LCO.

Consequently, we have found that there is a phase-relationship among the dominant coherent structures of the flow velocity field, the acoustic pressure field and the heat release rate field. The flame surface structures are high correlated with the structures in heat release rate. However, these structures contribute less to the coherent structures of flow velocity field. Temporal analysis exhibits that the subsystems of the combustor demonstrate the 1:1 and 2:1 frequency locking behaviour.

\section{Conclusion}
\label{sec:Conclusion}

In this paper, we present a framework for analyzing the synchronization characteristics of multiple flow parameters that are acquired simultaneously. This framework utilizes the extended proper orthogonal decomposition to look for correlations between coherent structures measured observed in planar velocity field measurements, and parameters such such heat-release rate, acoustic pressure fluctuations and OH-distribution. We characterize the coherent structures in the flow field using the proper orthogonal decomposition (POD). We also describe the structures in the OH-PLIF images and heat release rate distribution field using extended POD during the occurrence of various dynamical states of chaotic oscillations, thermoacoustic instability with period-1 and period-2 limit cycle oscillations. These dynamical states are observed with the variation of the amount of hydrogen in the fuel by keeping the thermal power and equivalence ratio constant. 

During the state of chaotic oscillations, we hardly find correlations among the structures in the flow, the flame and heat release rate distribution fields. However, during the state of P1 LCO, we observe that the structures with high energy coherent structures correlate with the high energy structures of the flame surface and heat release rate distribution fields as the structures occur at the same spatial regions. During the P2 LCO, the structures from the flame surface are strongly correlated with the structures of the heat release rate field. However, the structures from these fields contribute less to the coherent structures of the velocity field.

The time series analysis shows that the temporal coefficients of the first two modes dominate and decides the nature of the flow field during thermoacoustic instability with period-1 and period-2 limit cycle oscillations. Moreover, the patterns in the phase portrait of different pairs of the signals demonstrates 1:1 frequency locking behaviour among the acoustic field, heat release rate and the flow field during period-1 limit cycle oscillations. In case of period-2 limit cycle oscillations, we see both 1:1 and 2:1 frequency locking behaviour. 

Thus, with the help of temporal coefficients of dominating POD modes of the flow field, we provide a framework to understand the synchronization of the flow velocity field with the acoustic pressure and the heat release rate. Apart from this, we also find the correlations of the coherent structures in a flow field with the flame and heat release rate distribution field using the EPOD modes of OH-PLIF images and OH*-chemiluminescence fields. This study enhances the fundamental understanding of the coherence in the flow velocity and heat release rate distributions leading to different types of thermoacoustic instabilities which can occur in practical gas turbines.

%\appendix

 \vspace{0.2 cm}

\noindent
 \textbf{Acknowledgements:} Abhishek Kushwaha would like to acknowledge the Ministry of Human Resource Development, India for providing Ph.D scholarship. The authors would like to acknowledge Dr. Samadhan A. Pawar (UTIAS, Canada) for helpful discussions on this topic.

 \vspace{0.2 cm}
\noindent
 \textbf{Funding:} This project has received funding from the European Research Council (ERC) under the European Union’s Horizon 2020 research and innovation program (Grant Agreement No. 682383). This research work was also supported by the IoE initiative (SB/2021/0845/AE/MHRD/002696), IIT Madras, India.

 \vspace{0.2 cm}
 \noindent
 \textbf{Declaration of interests:} The authors report no conflict of interest.

 \vspace{0.2 cm}
 \noindent
 \textbf{Data availability statement:} The data that support the findings of this study are available upon reasonable request from the corresponding authors.

%\newpage

\bibliographystyle{jfm}
\bibliography{main}

\begin{thebibliography}{90}
\expandafter\ifx\csname natexlab\endcsname\relax\def\natexlab#1{#1}\fi
\def\au#1{#1} \def\ed#1{#1} \def\yr#1{#1}\def\at#1{#1}\def\jt#1{\textit{#1}}
  \def\bt#1{#1}\def\bvol#1{\textbf{#1}} \def\vol#1{#1} \def\pg#1{#1}
  \def\publ#1{#1}\def\arxiv#1{#1}\def\org#1{#1}\def\st#1{\textit{#1}}

\bibitem[{\AE}s{\o}y {\em et~al.\/}(2020){\AE}s{\o}y, Aguilar, Wiseman,
  Bothien, Worth \& Dawson]{aesoy2020scaling}
{\sc \au{{\AE}s{\o}y, E.}, \au{Aguilar, J.~G.}, \au{Wiseman, S.}, \au{Bothien,
  M.~R.}, \au{Worth, N.~A.} \& \au{Dawson, J.~R.}} \yr{2020}  \at{Scaling and
  prediction of transfer functions in lean premixed {H2/CH4}-flames}.
  \jt{Combust. Flame}  \bvol{215},  \pg{269--282}.

\bibitem[Ashwin {\em et~al.\/}(2001)Ashwin, Melbourne \&
  Nicol]{ashwin2001hypermeander}
{\sc \au{Ashwin, P.}, \au{Melbourne, I.} \& \au{Nicol, M.}} \yr{2001}
  \at{Hypermeander of spirals: local bifurcations and statistical properties}.
  \jt{Physica D.}  \bvol{156}~(3-4),  \pg{364--382}.

\bibitem[Berkooz {\em et~al.\/}(1993)Berkooz, Holmes \&
  Lumley]{berkooz1993proper}
{\sc \au{Berkooz, G.}, \au{Holmes, P.} \& \au{Lumley, J.~L.}} \yr{1993}
  \at{The proper orthogonal decomposition in the analysis of turbulent flows}.
  \jt{Annu. Rev. Fluid Mech.}  \bvol{25}~(1),  \pg{539--575}.

\bibitem[Birol(2019)]{birol2019future}
{\sc \au{Birol, F.}} \yr{2019}  \at{The future of hydrogen: Seizing today's
  opportunities}.  \jt{Report prepared by the IEA for the G20, 82-83, Japan} .

\bibitem[Bor{\'e}e(2003)]{boree2003extended}
{\sc \au{Bor{\'e}e, J.}} \yr{2003}  \at{Extended proper orthogonal
  decomposition: a tool to analyse correlated events in turbulent flows}.
  \jt{Exp. Fluids}  \bvol{35}~(2),  \pg{188--192}, {DOI:
  10.1007/s00348-003-0656-3}.

\bibitem[Boxx {\em et~al.\/}(2010)Boxx, St{\"o}hr, Carter \&
  Meier]{boxx2010temporally}
{\sc \au{Boxx, I.}, \au{St{\"o}hr, M.}, \au{Carter, C.} \& \au{Meier, W.}}
  \yr{2010}  \at{Temporally resolved planar measurements of transient phenomena
  in a partially pre-mixed swirl flame in a gas turbine model combustor}.
  \jt{Combust. Flame}  \bvol{157}~(8),  \pg{1510--1525}.

\bibitem[Candel {\em et~al.\/}(2009)Candel, Durox, Ducruix, Birbaud, Noiray \&
  Schuller]{candel2009flame}
{\sc \au{Candel, S.}, \au{Durox, D.}, \au{Ducruix, S.}, \au{Birbaud, A.-L.},
  \au{Noiray, N.} \& \au{Schuller, T.}} \yr{2009}  \at{Flame dynamics and
  combustion noise: progress and challenges}.  \jt{International J.
  Aeroacoustics}  \bvol{8}~(1),  \pg{1--56}.

\bibitem[Candel {\em et~al.\/}(2014)Candel, Durox, Schuller, Bourgouin \&
  Moeck]{candel2014dynamics}
{\sc \au{Candel, S.}, \au{Durox, D.}, \au{Schuller, T.}, \au{Bourgouin, J.-F.}
  \& \au{Moeck, J.~P.}} \yr{2014}  \at{Dynamics of swirling flames}.  \jt{Annu.
  Rev. Fluid Mech.}  \bvol{46},  \pg{147--173}, dOI:
  10.1146/annurev-fluid-010313-141300.

\bibitem[Chakravarthy {\em et~al.\/}(2007)Chakravarthy, Sivakumar \&
  Shreenivasan]{chakravarthy2007vortex}
{\sc \au{Chakravarthy, S.}, \au{Sivakumar, R.} \& \au{Shreenivasan, O.}}
  \yr{2007}  \at{Vortex-acoustic lock-on in bluff-body and backward-facing step
  combustors}.  \jt{Sadhana}  \bvol{32}~(1),  \pg{145--154},
  {DOI:10.1007/s12046-007-0013-y}.

\bibitem[Choi {\em et~al.\/}(2007)Choi, Rusak \& Kapila]{choi2007numerical}
{\sc \au{Choi, J.}, \au{Rusak, Z.} \& \au{Kapila, A.}} \yr{2007}  \at{Numerical
  simulation of premixed chemical reactions with swirl}.  \jt{Combust. Theor.
  Model.}  \bvol{11}~(6),  \pg{863--887}, {DOI:
  https://doi.org/10.1080/13647830701256085}.

\bibitem[Chterev \& Boxx(2019)]{chterevflame}
{\sc \au{Chterev, I.} \& \au{Boxx, I.}} \yr{2019} Flame topology and combustion
  instability limits of lean premixed hydrogen enriched flames.  \bt{In {\em
  27th International Colloquium on the Dynamics of Explosions and Reactive
  Systems, July 28 - August 2, Beijing, China\/}}. DOI:
  https://elib.dlr.de/130096/.

\bibitem[Chterev \& Boxx(2021)]{chterev2021effect}
{\sc \au{Chterev, I.} \& \au{Boxx, I.}} \yr{2021}  \at{Effect of hydrogen
  enrichment on the dynamics of a lean technically premixed elevated pressure
  flame}.  \jt{Combust. Flame}  \bvol{225},  \pg{149--159}, {DOI:
  10.1016/j.combustflame.2020.10.033}.

\bibitem[Chu(1965)]{chu1965energy}
{\sc \au{Chu, B.~T.}} \yr{1965}  \at{On the energy transfer to small
  disturbances in fluid flow (part i)}.  \jt{Acta Mech.}  \bvol{1}~(3),
  \pg{215--234}, {DOI: 10.1007/BF01387235}.

\bibitem[Cozzi \& Coghe(2006)]{cozzi2006behavior}
{\sc \au{Cozzi, F.} \& \au{Coghe, A.}} \yr{2006}  \at{Behavior of
  hydrogen-enriched non-premixed swirled natural gas flames}.  \jt{Int. J.
  Hydrogen Energy}  \bvol{31}~(6),  \pg{669--677}, {DOI:
  10.1016/j.ijhydene.2005.05.013}.

\bibitem[Davis {\em et~al.\/}(2013)Davis, Therkelsen, Littlejohn \&
  Cheng]{davis2013effects}
{\sc \au{Davis, D.}, \au{Therkelsen, P.}, \au{Littlejohn, D.} \& \au{Cheng,
  R.}} \yr{2013}  \at{Effects of hydrogen on the thermo-acoustics coupling
  mechanisms of low-swirl injector flames in a model gas turbine combustor}.
  \jt{Proc. Combust. Inst.}  \bvol{34}~(2),  \pg{3135--3143}, {DOI:
  10.1016/j.proci.2012.05.050}.

\bibitem[Dowling(1997)]{dowling1997nonlinear}
{\sc \au{Dowling, A.~P.}} \yr{1997}  \at{Nonlinear self-excited oscillations of
  a ducted flame}.  \jt{J. Fluid Mech.}  \bvol{346},  \pg{271--290}, {DOI:
  10.1017/S0022112097006484}.

\bibitem[Duwig \& Iudiciani(2010)]{duwig2010extended}
{\sc \au{Duwig, C.} \& \au{Iudiciani, P.}} \yr{2010}  \at{Extended proper
  orthogonal decomposition for analysis of unsteady flames}.  \jt{Flow,
  turbulence and combustion}  \bvol{84}~(1),  \pg{25--47}, {DOI:
  https://doi.org/10.1007/s10494-009-9210-6}.

\bibitem[Emadi {\em et~al.\/}(2012)Emadi, Karkow, Salameh, Gohil \&
  Ratner]{emadi2012flame}
{\sc \au{Emadi, M.}, \au{Karkow, D.}, \au{Salameh, T.}, \au{Gohil, A.} \&
  \au{Ratner, A.}} \yr{2012}  \at{Flame structure changes resulting from
  hydrogen-enrichment and pressurization for low-swirl premixed methane--air
  flames}.  \jt{Int. J. Hydrogen Energy}  \bvol{37}~(13),  \pg{10397--10404},
  {DOI: 10.1016/j.ijhydene.2012.04.017}.

\bibitem[Figura {\em et~al.\/}(2007)Figura, Lee, Quay \&
  Santavicca]{figura2007effects}
{\sc \au{Figura, L.}, \au{Lee, J.~G.}, \au{Quay, B.~D.} \& \au{Santavicca,
  D.~A.}} \yr{2007} The effects of fuel composition on flame structure and
  combustion dynamics in a lean premixed combustor.  \bt{In {\em Turbo Expo:
  Power for Land, Sea, and Air\/}}, ,  \vol{vol. 47918},  \pg{pp. 181--187}.
  {10.1115/GT2007-27298}.

\bibitem[Freitag \& Janicka(2007)]{freitag2007investigation}
{\sc \au{Freitag, M.} \& \au{Janicka, J.}} \yr{2007}  \at{Investigation of a
  strongly swirled unconfined premixed flame using {LES}}.  \jt{Proc. Combust.
  Inst.}  \bvol{31}~(1),  \pg{1477--1485}, {DOI: 10.1016/j.proci.2006.07.225}.

\bibitem[Gallaire {\em et~al.\/}(2006)Gallaire, Ruith, Meiburg, Chomaz \&
  Huerre]{gallaire2006spiral}
{\sc \au{Gallaire, F.}, \au{Ruith, M.}, \au{Meiburg, E.}, \au{Chomaz, J.-M.} \&
  \au{Huerre, P.}} \yr{2006}  \at{Spiral vortex breakdown as a global mode}.
  \jt{J. Fluid Mech.}  \bvol{549},  \pg{71--80}, {DOI}:
  10.1017/S0022112005007834.

\bibitem[Garc{\'\i}a-Armingol {\em et~al.\/}(2014)Garc{\'\i}a-Armingol,
  Hardalupas, Taylor \& Ballester]{garcia2014effect}
{\sc \au{Garc{\'\i}a-Armingol, T.}, \au{Hardalupas, Y.}, \au{Taylor, A.} \&
  \au{Ballester, J.}} \yr{2014}  \at{Effect of local flame properties on
  chemiluminescence-based stoichiometry measurement}.  \jt{Exp. Therm Fluid
  Sci.}  \bvol{53},  \pg{93--103}, {DOI: 10.1016/j.expthermflusci.2013.11.009}.

\bibitem[George {\em et~al.\/}(2018)George, Unni, Raghunathan \&
  Sujith]{george2018pattern}
{\sc \au{George, N.~B.}, \au{Unni, V.~R.}, \au{Raghunathan, M.} \& \au{Sujith,
  R.}} \yr{2018}  \at{Pattern formation during transition from combustion noise
  to thermoacoustic instability via intermittency}.  \jt{J. Fluid Mech.}
  \bvol{849},  \pg{615--644}, {DOI: 10.1017/jfm.2018.427}.

\bibitem[Gottwald \& Melbourne(2004)]{gottwald2004new}
{\sc \au{Gottwald, G.~A.} \& \au{Melbourne, I.}} \yr{2004}  \at{A new test for
  chaos in deterministic systems}.  \jt{Proc. R. Soc. London, Ser. A:
  Mathematical, Physical and Engineering Sciences}  \bvol{460}~(2042),
  \pg{603--611}, {DOI}: 10.1098/rspa.2003.1183.

\bibitem[Gottwald \& Melbourne(2009)]{gottwald2009implementation}
{\sc \au{Gottwald, G.~A.} \& \au{Melbourne, I.}} \yr{2009}  \at{On the
  implementation of the 0--1 test for chaos}.  \jt{SIAM Journal on Applied
  Dynamical Systems}  \bvol{8}~(1),  \pg{129--145}.

\bibitem[Greenslade~Jr(1993)]{greenslade1993all}
{\sc \au{Greenslade~Jr, T.~B.}} \yr{1993}  \at{All about {L}issajous figures}.
  \jt{Phys. Teach}  \bvol{31}~(6),  \pg{364--370}.

\bibitem[Guo {\em et~al.\/}(2010)Guo, Tayebi, Galizzi \&
  Escudie]{guo2010burning}
{\sc \au{Guo, H.}, \au{Tayebi, B.}, \au{Galizzi, C.} \& \au{Escudie, D.}}
  \yr{2010}  \at{Burning rates and surface characteristics of hydrogen-enriched
  turbulent lean premixed methane--air flames}.  \jt{Int. J. Hydrogen Energy}
  \bvol{35}~(20),  \pg{11342--11348}, {DOI: 10.1016/j.ijhydene.2010.07.066}.

\bibitem[Guo {\em et~al.\/}(2020)Guo, Wang, Zhang, Zhang \&
  Huang]{guo2020effect}
{\sc \au{Guo, S.}, \au{Wang, J.}, \au{Zhang, W.}, \au{Zhang, M.} \& \au{Huang,
  Z.}} \yr{2020}  \at{Effect of hydrogen enrichment on swirl/bluff-body lean
  premixed flame stabilization}.  \jt{Int. J. Hydrogen Energy}  \bvol{45}~(18),
   \pg{10906--10919}, {DOI: 10.1016/j.ijhydene.2020.02.020}.

\bibitem[Gupta {\em et~al.\/}(1984)Gupta, Lilley \& Syred]{gupta1984swirl}
{\sc \au{Gupta, A.~K.}, \au{Lilley, D.~G.} \& \au{Syred, N.}} \yr{1984}
  \at{Swirl flows}.  \jt{Tunbridge Wells} .

\bibitem[Halter {\em et~al.\/}(2007)Halter, Chauveau \&
  G{\"o}kalp]{halter2007characterization}
{\sc \au{Halter, F.}, \au{Chauveau, C.} \& \au{G{\"o}kalp, I.}} \yr{2007}
  \at{Characterization of the effects of hydrogen addition in premixed
  methane/air flames}.  \jt{Int. J. Hydrogen Energy}  \bvol{32}~(13),
  \pg{2585--2592}, {DOI: 10.1016/j.ijhydene.2006.11.033}.

\bibitem[Harvey(1962)]{harvey1962some}
{\sc \au{Harvey, J.}} \yr{1962}  \at{Some observations of the vortex breakdown
  phenomenon}.  \jt{Journal of Fluid Mechanics}  \bvol{14}~(4),  \pg{585--592}.

\bibitem[Hawkes \& Chen(2004)]{hawkes2004direct}
{\sc \au{Hawkes, E.~R.} \& \au{Chen, J.~H.}} \yr{2004}  \at{Direct numerical
  simulation of hydrogen-enriched lean premixed methane--air flames}.
  \jt{Combust. Flame}  \bvol{138}~(3),  \pg{242--258}, {DOI:
  10.1016/j.combustflame.2004.04.010}.

\bibitem[Hoffmann {\em et~al.\/}(1994)Hoffmann, Habisreuther \&
  Lenze]{hoffmann1994development}
{\sc \au{Hoffmann, S.}, \au{Habisreuther, P.} \& \au{Lenze, B.}} \yr{1994}
  \at{Development and assessment of correlations for predicting stability
  limits of swirling flames}.  \jt{Chem. Eng. Process. Process Intensif.}
  \bvol{33}~(5),  \pg{393--400}, {DOI: 10.1016/0255-2701(94)02011-6}.

\bibitem[Hong {\em et~al.\/}(2013)Hong, Shanbhogue, Speth \&
  Ghoniem]{hong2013phase}
{\sc \au{Hong, S.}, \au{Shanbhogue, S.~J.}, \au{Speth, R.~L.} \& \au{Ghoniem,
  A.~F.}} \yr{2013}  \at{On the phase between pressure and heat release
  fluctuations for propane/hydrogen flames and its role in mode transitions}.
  \jt{Combust. Flame}  \bvol{160}~(12),  \pg{2827--2842}, {DOI:
  10.1016/j.combustflame.2013.07.001}.

\bibitem[Hord(1978)]{hord1978hydrogen}
{\sc \au{Hord, J.}} \yr{1978}  \at{Is hydrogen a safe fuel?}  \jt{Int. J.
  Hydrogen Energy}  \bvol{3}~(2),  \pg{157--176}, {DOI:
  10.1016/0360-3199(78)90016-2}.

\bibitem[Huang \& Yang(2009)]{huang2009dynamics}
{\sc \au{Huang, Y.} \& \au{Yang, V.}} \yr{2009}  \at{Dynamics and stability of
  lean-premixed swirl-stabilized combustion}.  \jt{Prog. Energy Combust. Sci.}
  \bvol{35}~(4),  \pg{293--364}.

\bibitem[Janus {\em et~al.\/}(1997)Janus, Richards, Yip \&
  Robey]{janus1997effects}
{\sc \au{Janus, M.~C.}, \au{Richards, G.~A.}, \au{Yip, M.~J.} \& \au{Robey,
  E.~H.}} \yr{1997} Effects of ambient conditions and fuel composition on
  combustion stability.  \bt{In {\em Turbo Expo: Power for Land, Sea, and
  Air\/}}, ,  \vol{vol. 78699},  \pg{p. V002T06A035}. American Society of
  Mechanical Engineers, {DOI: 10.1115/97-GT-266}.

\bibitem[Juniper \& Sujith(2018)]{juniper2018sensitivity}
{\sc \au{Juniper, M.~P.} \& \au{Sujith, R.~I.}} \yr{2018}  \at{Sensitivity and
  nonlinearity of thermoacoustic oscillations}.  \jt{Annu. Rev. Fluid Mech.}
  \bvol{50},  \pg{661--689},
  {DOI}:https://doi.org/10.1146/annurev-fluid-122316-045125.

\bibitem[Kim {\em et~al.\/}(2009)Kim, Arghode, Linck \& Gupta]{kim2009hydrogen}
{\sc \au{Kim, H.~S.}, \au{Arghode, V.~K.}, \au{Linck, M.~B.} \& \au{Gupta,
  A.~K.}} \yr{2009}  \at{Hydrogen addition effects in a confined
  swirl-stabilized methane-air flame}.  \jt{Int. J. Hydrogen Energy}
  \bvol{34}~(2),  \pg{1054--1062}, {DOI}: 10.1016/j.ijhydene.2008.10.034.

\bibitem[Kushwaha {\em et~al.\/}(2021)Kushwaha, Kasthuri, Pawar, Sujith,
  Chterev \& Boxx]{kushwaha2021dynamical}
{\sc \au{Kushwaha, A.}, \au{Kasthuri, P.}, \au{Pawar, S.~A.}, \au{Sujith,
  R.~I.}, \au{Chterev, I.} \& \au{Boxx, I.}} \yr{2021}  \at{Dynamical
  characterization of thermoacoustic oscillations in a hydrogen-enriched
  partially premixed swirl-stabilized methane/air combustor}.  \jt{J. Eng. Gas
  Turbines Power}  \bvol{143}~(12).

\bibitem[Lee \& Kim(2020)]{lee2020combustion}
{\sc \au{Lee, T.} \& \au{Kim, K.~T.}} \yr{2020}  \at{Combustion dynamics of
  lean fully-premixed hydrogen-air flames in a mesoscale multinozzle array}.
  \jt{Combust. Flame}  \bvol{218},  \pg{234--246}.

\bibitem[Liang \& Maxworthy(2005)]{liang2005experimental}
{\sc \au{Liang, H.} \& \au{Maxworthy, T.}} \yr{2005}  \at{An experimental
  investigation of swirling jets}.  \jt{J. Fluid Mech.}  \bvol{525},
  \pg{115--159}, {DOI: https://doi.org/10.1017/S0022112004002629}.

\bibitem[Lieuwen(2021)]{lieuwen2021unsteady}
{\sc \au{Lieuwen, T.~C.}} \yr{2021} {\em Unsteady combustor physics\/}.
  \publ{Cambridge University Press}.

\bibitem[Lieuwen \& Yang(2005)]{lieuwen2005combustion}
{\sc \au{Lieuwen, T.~C.} \& \au{Yang, V.}} \yr{2005} {\em Combustion
  instabilities in gas turbine engines: operational experience, fundamental
  mechanisms, and modeling\/}.  \publ{American Institute of Aeronautics and
  Astronautics}.

\bibitem[Lohrasbi {\em et~al.\/}(2021)Lohrasbi, Hammer, Essl, Reiss, Defregger
  \& Sanz]{lohrasbi2021modification}
{\sc \au{Lohrasbi, S.}, \au{Hammer, R.}, \au{Essl, W.}, \au{Reiss, G.},
  \au{Defregger, S.} \& \au{Sanz, W.}} \yr{2021}  \at{A modification to
  extended proper orthogonal decomposition-based correlation analysis: The
  spatial consideration}.  \jt{Int. J. Heat Mass Transfer}  \bvol{175},
  \pg{121065}, {DOI: 10.1016/j.ijheatmasstransfer.2021.121065}.

\bibitem[Lumley(1967)]{lumley1967structure}
{\sc \au{Lumley, J.~L.}} \yr{1967}  \at{The structure of inhomogeneous
  turbulent flows}.  \jt{Atmospheric turbulence and radio wave propagation}
  \pg{pp. 166--178}.

\bibitem[Mandilas {\em et~al.\/}(2007)Mandilas, Ormsby, Sheppard \&
  Woolley]{mandilas2007effects}
{\sc \au{Mandilas, C.}, \au{Ormsby, M.}, \au{Sheppard, C.} \& \au{Woolley, R.}}
  \yr{2007}  \at{Effects of hydrogen addition on laminar and turbulent premixed
  methane and iso-octane--air flames}.  \jt{Proc. Combust. Inst.}
  \bvol{31}~(1),  \pg{1443--1450}, {DOI: 10.1016/j.proci.2006.07.157}.

\bibitem[Moeck \& Paschereit(2012)]{moeck2012nonlinear}
{\sc \au{Moeck, J.~P.} \& \au{Paschereit, C.~O.}} \yr{2012}  \at{Nonlinear
  interactions of multiple linearly unstable thermoacoustic modes}.  \jt{Int.
  J. Spray Combust. Dyn.}  \bvol{4}~(1),  \pg{1--27}.

\bibitem[Mondal {\em et~al.\/}(2017)Mondal, Pawar \&
  Sujith]{mondal2017synchronous}
{\sc \au{Mondal, S.}, \au{Pawar, S.} \& \au{Sujith, R.~I.}} \yr{2017}
  \at{Synchronous behaviour of two interacting oscillatory systems undergoing
  quasiperiodic route to chaos}.  \jt{Chaos: An Interdisciplinary Journal of
  Nonlinear Science}  \bvol{27}~(10),  \pg{103119}, dOI: 10.1063/1.4991744.

\bibitem[Nair \& Sujith(2014)]{nair2014multifractality}
{\sc \au{Nair, V.} \& \au{Sujith, R.~I.}} \yr{2014}  \at{Multifractality in
  combustion noise: predicting an impending combustion instability}.  \jt{J.
  Fluid Mech.}  \bvol{747},  \pg{635--655}, dOI: 10.1017/jfm.2014.171.

\bibitem[Nair {\em et~al.\/}(2013)Nair, Thampi, Karuppusamy, Gopalan \&
  Sujith]{nair2013loss}
{\sc \au{Nair, V.}, \au{Thampi, G.}, \au{Karuppusamy, S.}, \au{Gopalan, S.} \&
  \au{Sujith, R.}} \yr{2013}  \at{Loss of chaos in combustion noise as a
  precursor of impending combustion instability}.  \jt{Int. J. Spray Combust.
  Dyn.}  \bvol{5}~(4),  \pg{273--290},
  dOI:https://doi.org/10.1260/1756-8277.5.4.273.

\bibitem[Nakahara \& Kido(2008)]{nakahara2008study}
{\sc \au{Nakahara, M.} \& \au{Kido, H.}} \yr{2008}  \at{Study on the turbulent
  burning velocity of hydrogen mixtures including hydrocarbon}.  \jt{AIAA J.}
  \bvol{46}~(7),  \pg{1569--1575}, {DOI}: 10.2514/1.23560.

\bibitem[Nam {\em et~al.\/}(2019)Nam, Lee, Joo, Yoon \& Yoh]{nam2019numerical}
{\sc \au{Nam, J.}, \au{Lee, Y.}, \au{Joo, S.}, \au{Yoon, Y.} \& \au{Yoh,
  J.~J.}} \yr{2019}  \at{Numerical analysis of the effect of the hydrogen
  composition on a partially premixed gas turbine combustor}.  \jt{Int. J.
  Hydrogen Energy}  \bvol{44}~(12),  \pg{6278--6286}, {DOI}:
  https://doi.org/10.1016/j.ijhydene.2019.01.066.

\bibitem[Oberleithner {\em et~al.\/}(2011)Oberleithner, Sieber, Nayeri,
  Paschereit, Petz, Hege, Noack \& Wygnanski]{oberleithner2011three}
{\sc \au{Oberleithner, K.}, \au{Sieber, M.}, \au{Nayeri, C.~N.},
  \au{Paschereit, C.~O.}, \au{Petz, C.}, \au{Hege, H.-C.}, \au{Noack, B.~R.} \&
  \au{Wygnanski, I.}} \yr{2011}  \at{Three-dimensional coherent structures in a
  swirling jet undergoing vortex breakdown: stability analysis and empirical
  mode construction}.  \jt{J. Fluid Mech.}  \bvol{679},  \pg{383--414}, {DOI}:
  https://doi.org/10.1017/jfm.2011.141.

\bibitem[Oberleithner {\em et~al.\/}(2015)Oberleithner, St{\"o}hr, Im, Arndt \&
  Steinberg]{oberleithner2015formation}
{\sc \au{Oberleithner, K.}, \au{St{\"o}hr, M.}, \au{Im, S.~H.}, \au{Arndt,
  C.~M.} \& \au{Steinberg, A.~M.}} \yr{2015}  \at{Formation and flame-induced
  suppression of the precessing vortex core in a swirl combustor: experiments
  and linear stability analysis}.  \jt{Combust. Flame}  \bvol{162}~(8),
  \pg{3100--3114}, {DOI: 10.1016/j.combustflame.2015.02.015 }.

\bibitem[Oberleithner {\em et~al.\/}(2013)Oberleithner, Terhaar, Rukes \&
  Oliver~Paschereit]{oberleithner2013nonuniform}
{\sc \au{Oberleithner, K.}, \au{Terhaar, S.}, \au{Rukes, L.} \&
  \au{Oliver~Paschereit, C.}} \yr{2013}  \at{Why nonuniform density suppresses
  the precessing vortex core}.  \jt{J. Engng. Gas Turbines Pow.}
  \bvol{135}~(12).

\bibitem[Pawar {\em et~al.\/}(2017)Pawar, Seshadri, Unni \&
  Sujith]{pawar2017thermoacoustic}
{\sc \au{Pawar, S.~A.}, \au{Seshadri, A.}, \au{Unni, V.~R.} \& \au{Sujith,
  R.~I.}} \yr{2017}  \at{Thermoacoustic instability as mutual synchronization
  between the acoustic field of the confinement and turbulent reactive flow}.
  \jt{J. Fluid Mech.}  \bvol{827},  \pg{664--693}, {DOI}: 10.1017/jfm.2017.438.

\bibitem[Pignatelli {\em et~al.\/}(2022)Pignatelli, Kim, Subash, Liu, Szasz,
  Bai, Brackmann, Ald{\'e}n \& L{\"o}rstad]{pignatelli2022pilot}
{\sc \au{Pignatelli, F.}, \au{Kim, H.}, \au{Subash, A.}, \au{Liu, X.},
  \au{Szasz, R.}, \au{Bai, X.}, \au{Brackmann, C.}, \au{Ald{\'e}n, M.} \&
  \au{L{\"o}rstad, D.}} \yr{2022}  \at{Pilot impact on turbulent premixed
  methane/air and hydrogen-enriched methane/air flames in a laboratory-scale
  gas turbine model combustor}.  \jt{Int. J. Hydrogen Energy}  \bvol{47}~(60),
  \pg{25404--25417}, {DOI}: https://doi.org/10.1016/j.ijhydene.2022.05.282.

\bibitem[Pikovsky {\em et~al.\/}(2001)Pikovsky, Rosenblum, Kurths {\em
  et~al.\/}]{pikovsky2001universal}
{\sc \au{Pikovsky, A.}, \au{Rosenblum, M.}, \au{Kurths, J.} \& \au{others}}
  \yr{2001}  \at{A universal concept in nonlinear sciences}.  \jt{Self}
  \bvol{2},  \pg{3}.

\bibitem[Poinsot {\em et~al.\/}(1987)Poinsot, Trouve, Veynante, Candel \&
  Esposito]{poinsot1987vortex}
{\sc \au{Poinsot, T.~J.}, \au{Trouve, A.~C.}, \au{Veynante, D.~P.}, \au{Candel,
  S.~M.} \& \au{Esposito, E.~J.}} \yr{1987}  \at{Vortex-driven acoustically
  coupled combustion instabilities}.  \jt{J. Fluid Mech.}  \bvol{177},
  \pg{265--292}, {DOI}: https://doi.org/10.1017/S0022112087000958.

\bibitem[Putnam(1971)]{putnam1971combustion}
{\sc \au{Putnam, A.~A.}} \yr{1971} {\em Combustion-driven oscillations in
  industry\/}.  \publ{Elsevier Publishing Company}.

\bibitem[Qadri {\em et~al.\/}(2013)Qadri, Mistry \&
  Juniper]{qadri2013structural}
{\sc \au{Qadri, U.~A.}, \au{Mistry, D.} \& \au{Juniper, M.~P.}} \yr{2013}
  \at{Structural sensitivity of spiral vortex breakdown}.  \jt{J. Fluid Mech.}
  \bvol{720},  \pg{558--581}, {DOI}: https://doi.org/10.1017/jfm.2013.34.

\bibitem[Rashwan {\em et~al.\/}(2016)Rashwan, Nemitallah \&
  Habib]{rashwan2016review}
{\sc \au{Rashwan, S.~S.}, \au{Nemitallah, M.~A.} \& \au{Habib, M.~A.}}
  \yr{2016}  \at{Review on premixed combustion technology: stability, emission
  control, applications, and numerical case study}.  \jt{Energy \& Fuels}
  \bvol{30}~(12),  \pg{9981--10014}, {DOI}:
  https://doi.org/10.1021/acs.energyfuels.6b02386.

\bibitem[Rayleigh(1878)]{rayleigh1878explanation}
{\sc \au{Rayleigh, J. W.~S.}} \yr{1878}  \at{The explanation of certain
  acoustical phenomena}.  \jt{Nature}  \bvol{18}~(455),  \pg{319--321},
  {DOI:10.1038/018319a0}.

\bibitem[Reddy {\em et~al.\/}(2006)Reddy, Sujith \&
  Chakravarthy]{reddy2006swirler}
{\sc \au{Reddy, A.~P.}, \au{Sujith, R.} \& \au{Chakravarthy, S.}} \yr{2006}
  \at{Swirler flow field characteristics in a sudden expansion combustor
  geometry}.  \jt{J. Propul. Power}  \bvol{22}~(4),  \pg{800--808}, {DOI:
  https://doi.org/10.2514/1.15940}.

\bibitem[Rosenblum {\em et~al.\/}(1996)Rosenblum, Pikovsky \&
  Kurths]{rosenblum1996phase}
{\sc \au{Rosenblum, M.~G.}, \au{Pikovsky, A.~S.} \& \au{Kurths, J.}} \yr{1996}
  \at{Phase synchronization of chaotic oscillators}.  \jt{Physical review
  letters}  \bvol{76}~(11),  \pg{1804}.

\bibitem[Schadow \& Gutmark(1992)]{schadow1992combustion}
{\sc \au{Schadow, K.} \& \au{Gutmark, E.}} \yr{1992}  \at{Combustion
  instability related to vortex shedding in dump combustors and their passive
  control}.  \jt{Prog. Energy Combust. Sci.}  \bvol{18}~(2),  \pg{117--132},
  {DOI}: https://doi.org/10.1016/0360-1285(92)90020-2.

\bibitem[Schefer(2003)]{schefer2003hydrogen}
{\sc \au{Schefer, R.}} \yr{2003}  \at{Hydrogen enrichment for improved lean
  flame stability}.  \jt{Int. J. Hydrogen Energy}  \bvol{28}~(10),
  \pg{1131--1141}, {DOI}: 10.1016/S0360-3199(02)00199-4.

\bibitem[Schefer {\em et~al.\/}(2002)Schefer, Wicksall \&
  Agrawal]{schefer2002combustion}
{\sc \au{Schefer, R.~W.}, \au{Wicksall, D.} \& \au{Agrawal, A.}} \yr{2002}
  \at{Combustion of hydrogen-enriched methane in a lean premixed
  swirl-stabilized burner}.  \jt{Proc. Combust. Inst.}  \bvol{29}~(1),
  \pg{843--851}, {DOI:10.1016/S1540-7489(02)80108-0}.

\bibitem[Schmitt {\em et~al.\/}(2007)Schmitt, Poinsot, Schuermans \&
  Geigle]{schmitt2007large}
{\sc \au{Schmitt, P.}, \au{Poinsot, T.}, \au{Schuermans, B.} \& \au{Geigle,
  K.-P.}} \yr{2007}  \at{Large-eddy simulation and experimental study of heat
  transfer, nitric oxide emissions and combustion instability in a swirled
  turbulent high-pressure burner}.  \jt{J. Fluid Mech.}  \bvol{570},
  \pg{17--46}, {DOI}: https://doi.org/10.1017/S0022112006003156.

\bibitem[Shanbhogue {\em et~al.\/}(2016)Shanbhogue, Sanusi, Taamallah, Habib,
  Mokheimer \& Ghoniem]{shanbhogue2016flame}
{\sc \au{Shanbhogue, S.~J.}, \au{Sanusi, Y.~S.}, \au{Taamallah, S.}, \au{Habib,
  M.~A.}, \au{Mokheimer, E. M.~A.} \& \au{Ghoniem, A.~F.}} \yr{2016}  \at{Flame
  macrostructures, combustion instability and extinction strain scaling in
  swirl-stabilized premixed {CH4/H2} combustion}.  \jt{Combust. Flame}
  \bvol{163},  \pg{494--507}, {DOI}:
  https://doi.org/10.1016/j.combustflame.2015.10.026.

\bibitem[Sieber {\em et~al.\/}(2017)Sieber, Oliver~Paschereit \&
  Oberleithner]{sieber2017advanced}
{\sc \au{Sieber, M.}, \au{Oliver~Paschereit, C.} \& \au{Oberleithner, K.}}
  \yr{2017}  \at{Advanced identification of coherent structures in
  swirl-stabilized combustors}.  \jt{J. Eng. Gas Turbines Power}
  \bvol{139}~(2), {DOI}: https://doi.org/10.1115/1.4034261.

\bibitem[Singh {\em et~al.\/}(2022)Singh, Roy, Dhadphale, Chaudhuri \&
  Sujith]{singh2022mean}
{\sc \au{Singh, S.}, \au{Roy, A.}, \au{Dhadphale, J.~M.}, \au{Chaudhuri, S.} \&
  \au{Sujith, R.~I.}} \yr{2022}  \at{Mean-field synchronization model of
  turbulent thermoacoustic transitions}.  \jt{arXiv preprint arXiv:2208.11550}
  .

\bibitem[Sirovich(1987)]{sirovich1987turbulence}
{\sc \au{Sirovich, L.}} \yr{1987}  \at{Turbulence and the dynamics of coherent
  structures. {I}. coherent structures}.  \jt{Q. Appl. Math.}  \bvol{45}~(3),
  \pg{561--571}, {DOI}: https://doi.org/10.1090/qam/910462.

\bibitem[Sterling \& Zukoski(1987)]{sterling1987longitudinal}
{\sc \au{Sterling, J.} \& \au{Zukoski, E.~E.}} \yr{1987}  \at{Longitudinal mode
  combustion instabilities in a dump combustor} .

\bibitem[St{\"o}hr {\em et~al.\/}(2011)St{\"o}hr, Sadanandan \&
  Meier]{stohr2011phase}
{\sc \au{St{\"o}hr, M.}, \au{Sadanandan, R.} \& \au{Meier, W.}} \yr{2011}
  \at{Phase-resolved characterization of vortex--flame interaction in a
  turbulent swirl flame}.  \jt{Exp. Fluids}  \bvol{51}~(4),  \pg{1153--1167}.

\bibitem[Strakey {\em et~al.\/}(2007)Strakey, Sidwell \&
  Ontko]{strakey2007investigation}
{\sc \au{Strakey, P.}, \au{Sidwell, T.} \& \au{Ontko, J.}} \yr{2007}
  \at{Investigation of the effects of hydrogen addition on lean extinction in a
  swirl stabilized combustor}.  \jt{Proc. Combust. Inst.}  \bvol{31}~(2),
  \pg{3173--3180}, {DOI}: https://doi.org/10.1016/j.proci.2006.07.077.

\bibitem[Sui {\em et~al.\/}(2017)Sui, Zhao, Zhang \& Gao]{sui2017experimental}
{\sc \au{Sui, J.}, \au{Zhao, D.}, \au{Zhang, B.} \& \au{Gao, N.}} \yr{2017}
  \at{Experimental study of {R}ijke-type thermoacoustic instability by using
  proper orthogonal decomposition method}.  \jt{Exp. Therm Fluid Sci.}
  \bvol{81},  \pg{336--344}, {DOI}:
  https://doi.org/10.1016/j.expthermflusci.2016.10.026.

\bibitem[Sujith \& Pawar(2021)]{sujith2021thermoacoustic}
{\sc \au{Sujith, R.~I.} \& \au{Pawar, S.~A.}} \yr{2021} {\em Thermoacoustic
  Instability: A Complex Systems Perspective\/}.  \publ{Springer Nature}.

\bibitem[Syred {\em et~al.\/}(2014)Syred, Giles, Lewis, Abdulsada, Medina,
  Marsh, Bowen \& Griffiths]{syred2014effect}
{\sc \au{Syred, N.}, \au{Giles, A.}, \au{Lewis, J.}, \au{Abdulsada, M.},
  \au{Medina, A.~V.}, \au{Marsh, R.}, \au{Bowen, P.~J.} \& \au{Griffiths,
  A.~J.}} \yr{2014}  \at{Effect of inlet and outlet configurations on blow-off
  and flashback with premixed combustion for methane and a high hydrogen
  content fuel in a generic swirl burner}.  \jt{Appl. Energy}  \bvol{116},
  \pg{288--296}, {DOI}: https://doi.org/10.1016/j.apenergy.2013.11.071.

\bibitem[Taamallah {\em et~al.\/}(2015)Taamallah, LaBry, Shanbhogue \&
  Ghoniem]{taamallah2015thermo}
{\sc \au{Taamallah, S.}, \au{LaBry, Z.~A.}, \au{Shanbhogue, S.~J.} \&
  \au{Ghoniem, A.~F.}} \yr{2015}  \at{Thermo-acoustic instabilities in lean
  premixed swirl-stabilized combustion and their link to acoustically coupled
  and decoupled flame macrostructures}.  \jt{Proc. Combust. Inst.}
  \bvol{35}~(3),  \pg{3273--3282}, {DOI}:
  https://doi.org/10.1016/j.proci.2014.07.002.

\bibitem[Taira {\em et~al.\/}(2017)Taira, Brunton, Dawson, Rowley, Colonius,
  McKeon, Schmidt, Gordeyev, Theofilis \& Ukeiley]{taira2017modal}
{\sc \au{Taira, K.}, \au{Brunton, S.~L.}, \au{Dawson, S.~T.}, \au{Rowley,
  C.~W.}, \au{Colonius, T.}, \au{McKeon, B.~J.}, \au{Schmidt, O.~T.},
  \au{Gordeyev, S.}, \au{Theofilis, V.} \& \au{Ukeiley, L.~S.}} \yr{2017}
  \at{Modal analysis of fluid flows: An overview}.  \jt{{AIAA J.}}
  \bvol{55}~(12),  \pg{4013--4041}, {DOI: https://doi.org/10.2514/1.J056060}.

\bibitem[Tuncer {\em et~al.\/}(2009)Tuncer, Acharya \& Uhm]{tuncer2009dynamics}
{\sc \au{Tuncer, O.}, \au{Acharya, S.} \& \au{Uhm, J.}} \yr{2009}
  \at{Dynamics, {NO}x and flashback characteristics of confined premixed
  hydrogen-enriched methane flames}.  \jt{Int. J. Hydrogen Energy}
  \bvol{34}~(1),  \pg{496--506}, {DOI}:
  https://doi.org/10.1016/j.ijhydene.2008.09.075.

\bibitem[Wang {\em et~al.\/}(2019)Wang, Ma \& Liu]{wang2019proper}
{\sc \au{Wang, P.}, \au{Ma, H.} \& \au{Liu, Y.}} \yr{2019}  \at{Proper
  orthogonal decomposition and extended-proper orthogonal decomposition
  analysis of pressure fluctuations and vortex structures inside a steam
  turbine control valve}.  \jt{J. Eng. Gas Turbines Power}  \bvol{141}~(4),
  {DOI}: https://doi.org/10.1115/1.4040903.

\bibitem[Wicksall {\em et~al.\/}(2005)Wicksall, Agrawal, Schefer \&
  Keller]{wicksall2005interaction}
{\sc \au{Wicksall, D.}, \au{Agrawal, A.}, \au{Schefer, R.} \& \au{Keller, J.}}
  \yr{2005}  \at{The interaction of flame and flow field in a lean premixed
  swirl-stabilized combustor operated on {H2/CH4/air}}.  \jt{Proc. Combust.
  Inst.}  \bvol{30}~(2),  \pg{2875--2883}, {DOI}:
  https://doi.org/10.1016/j.proci.2004.07.021.

\bibitem[Xia {\em et~al.\/}(2022)Xia, Han, Wei, Zhang, Wang, Huang \&
  Hasse]{xia2022numerical}
{\sc \au{Xia, H.}, \au{Han, W.}, \au{Wei, X.}, \au{Zhang, M.}, \au{Wang, J.},
  \au{Huang, Z.} \& \au{Hasse, C.}} \yr{2022}  \at{Numerical investigation of
  boundary layer flashback of {CH4/H2/air} swirl flames under different thermal
  boundary conditions in a bluff-body swirl burner}.  \jt{Proc. Combust. Inst.}
  {DOI}: https://doi.org/10.1016/j.proci.2022.07.040.

\bibitem[Zhang \& Ratner(2019)]{zhang2019experimental}
{\sc \au{Zhang, J.} \& \au{Ratner, A.}} \yr{2019}  \at{Experimental study on
  the excitation of thermoacoustic instability of hydrogen-methane/air premixed
  flames under atmospheric and elevated pressure conditions}.  \jt{Int. J.
  Hydrogen Energy}  \bvol{44}~(39),  \pg{21324--21335}, {DOI:
  10.1016/j.ijhydene.2019.06.142 }.

\bibitem[Zhang {\em et~al.\/}(2011)Zhang, Shanbhogue, Shreekrishna, Lieuwen \&
  O'Connor]{zhang2011strain}
{\sc \au{Zhang, Q.}, \au{Shanbhogue, S.~J.}, \au{Shreekrishna}, \au{Lieuwen,
  T.} \& \au{O'Connor, J.}} \yr{2011}  \at{Strain characteristics near the
  flame attachment point in a swirling flow}.  \jt{Combust. Sci. Technol.}
  \bvol{183}~(7),  \pg{665--685}, {DOI}:
  https://doi.org/10.1080/00102202.2010.537288.

\bibitem[Zhang {\em et~al.\/}(2020)Zhang, Wang, Lin, Mao, Xia, Zhang \&
  Huang]{zhang2020effect}
{\sc \au{Zhang, W.}, \au{Wang, J.}, \au{Lin, W.}, \au{Mao, R.}, \au{Xia, H.},
  \au{Zhang, M.} \& \au{Huang, Z.}} \yr{2020}  \at{Effect of differential
  diffusion on turbulent lean premixed hydrogen enriched flames through
  structure analysis}.  \jt{Int. J. Hydrogen Energy}  \bvol{45}~(18),
  \pg{10920--10931}, {DOI}: https://doi.org/10.1016/j.ijhydene.2020.02.032.

\bibitem[Zhen {\em et~al.\/}(2012)Zhen, Cheung, Leung \& Choy]{zhen2012effects}
{\sc \au{Zhen, H.}, \au{Cheung, C.~S.}, \au{Leung, C.~W.} \& \au{Choy, Y.~S.}}
  \yr{2012}  \at{Effects of hydrogen concentration on the emission and heat
  transfer of a premixed lpg-hydrogen flame}.  \jt{international journal of
  hydrogen energy}  \bvol{37}~(7),  \pg{6097--6105}.

\end{thebibliography}


\begin{thebibliography}{5}
\expandafter\ifx\csname natexlab\endcsname\relax\def\natexlab#1{#1}\fi
\def\au#1{#1} \def\ed#1{#1} \def\yr#1{#1}\def\at#1{#1}\def\jt#1{\textit{#1}}
  \def\bt#1{#1}\def\bvol#1{\textbf{#1}} \def\vol#1{#1} \def\pg#1{#1}
  \def\publ#1{#1}\def\arxiv#1{#1}\def\org#1{#1}\def\st#1{\textit{#1}}

\bibitem[Fouda {\em et~al.\/}(2014)Fouda, Bodo, Sabat \&
  Effa]{armand2014modified}
{\sc \au{Fouda, J. S. A.~E.}, \au{Bodo, B.}, \au{Sabat, S.~L.} \& \au{Effa,
  J.~Y.}} \yr{2014}  \at{A modified 0-1 test for chaos detection in oversampled
  time series observations}.  \jt{Int. J. Bifurcation Chaos}  \bvol{24}~(05),
  \pg{1450063}, dOI:https://doi.org/10.1142/S0218127414500631.

\bibitem[Kushwaha {\em et~al.\/}(2021)Kushwaha, Kasthuri, Pawar, Sujith,
  Chterev \& Boxx]{kushwaha2021dynamical}
{\sc \au{Kushwaha, A.}, \au{Kasthuri, P.}, \au{Pawar, S.~A.}, \au{Sujith,
  R.~I.}, \au{Chterev, I.} \& \au{Boxx, I.}} \yr{2021}  \at{Dynamical
  characterization of thermoacoustic oscillations in a hydrogen-enriched
  partially premixed swirl-stabilized methane/air combustor}.  \jt{J. Eng. Gas
  Turbines Power}  \bvol{143}~(12).

\bibitem[Melosik \& Marszalek(2016)]{melosik20160}
{\sc \au{Melosik, M.} \& \au{Marszalek, W.}} \yr{2016}  \at{On the 0/1 test for
  chaos in continuous systems}.  \jt{Bulletin of the Polish Academy of
  Sciences: Technical Sciences} ~(3),  \pg{521--528}.

\bibitem[Toker {\em et~al.\/}(2020)Toker, Sommer \&
  D’Esposito]{toker2020simple}
{\sc \au{Toker, D.}, \au{Sommer, F.~T.} \& \au{D’Esposito, M.}} \yr{2020}
  \at{A simple method for detecting chaos in nature}.  \jt{Commun. Biol.}
  \bvol{3}~(1),  \pg{1--13}, dOI:https://doi.org/10.1038/s42003-019-0715-9.

\bibitem[Wilke(1950)]{wilke1950viscosity}
{\sc \au{Wilke, C.}} \yr{1950}  \at{A viscosity equation for gas mixtures}.
  \jt{Chin. J. Chem. Phys. Chinese}  \bvol{18}~(4),  \pg{517--519}, {DOI}:
  https://doi.org/10.1063/1.1747673.

\end{thebibliography}

\end{document}

\maketitle

\vspace{0.2 cm}
%\setstretch{2}
This supplemental materials includes a comprehensive discussion on the experimental setup and the diagnostic techniques in \S\ref{Sec1-Expt}, sampling requirements of the $0-1$ chaos test in \S\ref{Sec2-Sampling}. In \S\ref{Sec3-axialcomponent EPOD}, the results pertaining to the streamwise component of the velocity field are presented and their relevance discussed.

\section{Experimental setup, instrumentation and diagnostic techniques}
\label{Sec1-Expt}
The experiments were performed at atmospheric pressure in the technically premixed gas turbine model combustor (PRECCINSTA configuration) shown in figure 1 of the manuscript. Air enters the cylindrical plenum before passing through a swirl generator with 12 radial swirl channels. Fuel (methane with variable quantities of hydrogen addition) is injected into each swirl channel through a 1 mm aperture. Through a conical nozzle with an exit diameter ($D$ = 27.85 mm), the partially premixed swirling reactant flow reaches the combustion chamber. The chamber has a height of 114 mm and a square cross-section area of 85 $\times$ 85 mm$^2$. Metal brackets at the corner are used for supporting quartz glass to give optical access to the combustion chamber. The exit of the combustor consists of a conical nozzle followed with an exhaust duct. The contraction ratio (ratio of the area of the exit of the nozzle at the inlet of the combustor to the cross-sectional area of the combustion chamber) is 0.17. This contraction ratio ensures the combustion at atmospheric pressure inside the combustion chamber.

Three separate mass flow sensing meters are used to control the mass flow rate of methane ($\dot{m}_{\rm {CH_4}}$), hydrogen ($\dot{m}_{\rm H_2}$) and air ($\dot{m}_{a}$). The addition of hydrogen with methane is done in such a way that the global equivalence ratio ($\phi_g$) and thermal power ($P_{th}$) remain constant for each case. These operating conditions are obtained by simultaneously decreasing $\dot{m}_{\rm CH_4}$ and increasing $\dot{m}_{\rm H_2}$. The thermal power for the combustor is measured using $P_{th} = \dot{m}_{\rm CH_4}h_{\rm CH_4} + \dot{m}_{\rm H_2}h_{\rm H_2}$, where $h_{\rm CH_4}$ and $h_{\rm H_2}$ are the calorific values of methane ($h_{\rm CH_4} = 55.5$ MJ/kg) and hydrogen ($h_{\rm H_2} = 141.7$ MJ/kg) respectively. The thermal power $P_{th}$ for the experiments varies from 22.17 kW to 25.31 kW as the volumetric percentage of $\rm H_2$ varies. The nominal Reynolds number $Re$ for the experiments is calculated using the expression,

\begin{equation}
    Re = \frac{4\dot{m}_t}{\mu \pi L_c}
\end{equation}
where $\dot{m}_t(= \dot{m}_{\rm CH_4}+\dot{m}_{\rm H_2}+\dot{m}_a)$ denotes the total mass flow rate in kg/s. The value of the dynamic viscosity ($\mu$) for the mixture of $\rm H_2-\rm CH_4-$air is computed according to \cite{wilke1950viscosity}. The diameter ($D$) of the nozzle at the inlet of the combustion chamber is used as the characteristic length ($L_c$). $Re$ ranges from 2.3 $\times 10^4$ to 2.8 $\times 10^4$. The global equivalence ratio ($\phi_g$), the thermal power ($P_{th}$) and the Reynolds number ($Re$) are listed in the table 1 in the main manuscript. 

Various measurements were performed to understand the state of the system. To measure acoustic pressure fluctuations in the combustion chamber associated with different dynamical states, amplitude and phase calibrated microphone probe with a Brüel and Kjaer (B$\&$K) Type 4939 ¼-inch Free-field condenser microphone was used. The sensitivity of the microphone is 4 mV/Pa and the uncertainty is $\pm$3 dB. The microphone was positioned 20 mm from the dump plane of the combustor and recorded data at a sampling frequency of 100 kHz. 

Stereoscopic particle image velocimetry (sPIV) is used to measure the flow characteristics inside the combustor. A dual-cavity, diode-pumped solid state laser (Edgewave, IS200-2-LD, up to 9 mJ/pulse at 532 nm) and a pair of CMOS cameras (Phantom v1212) positioned on the opposite sides of the laser sheet, looking down into the combustor, make up the stereo-PIV system. 10,000 (1 second) PIV images are acquired at a resolution of 640 $\times$ 800 pixels for each operational state. The PIV measurement domain spans a region of 65 $\times$ 50 mm$^2$, ($-32 < x < 32$ mm and $0 < y < 50$ mm in figure 1 of manuscript). The projected pixel resolution of the PIV system is 0.08 mm/pixel. The inter-frame pulse separation ($\Delta$t) is set to 10 ms. A pair of cylindrical lenses (\textit{f} = $-38$ mm and 250 mm) are used to shape the beam into a sheet, and a third cylindrical lens (\textit{f} = 700 mm) is used to thin it to a waist. Titanium dioxide (TiO$_2$) particles with a nominal diameter of 1 $\upmu$m is seeded into the air flow inside the combustor to obtain the scattered fluorescence. A commercial, multi-pass adaptive window offset cross-correlation technique is used to accomplish image mapping, calibration, and particle cross-correlations (LaVision DaVis 10). The final interrogation window size id 16 $\times$ 16 pixels with a 50\% overlap. This corresponds to a window size of 1.3 mm and vector spacing of 0.65 mm. In this manner, the three components of velocity in a plane aligned with the burner center-line are quantified. Based on the correlation statistics in LaVision DaVis, the uncertainty in the instantaneous velocities is estimated to be $\pm 0.7$ m/s for the in-plane components ($x$-$y$) and $\pm 1.8$ m/s for the out-of-plane component ($z$-axis). Figure 1b of manuscript shows a typical instance of the velocity field obtained from the PIV measurements.

An intensified high-speed CMOS camera (LaVision HSS 5 with LaVision HS-IRO) equipped with a fast UV objective lens (Halle, $f$ = 64 mm, $f$/2.0) and a bandpass filter (300–325 nm) is used to image the line of sight integrated chemiluminescence from the self-excited OH* radical over a 512 $\times$ 512 pixel resolution. Depending on the signal intensity, the intensifier gate time is ranged from 25 to 50 $\upmu$s. The number of images for each recording is limited by the onboard memory of the camera. For each condition, 8192 frames were captured. The global heat release rate $\dot{q}^\prime(t)$ signal is calculated by integrating the intensity of all pixels.

The OH-planar laser-induced fluorescence (PLIF) imaging system consists of a frequency-doubled dye laser pumped by a high-speed, pulsed Nd:YAG laser (Edgewave IS400-2-L, 150 W at 532 nm and 10 kHz) and an intensified high-speed CMOS camera system. The dye laser system (Sirah Credo) was operated in the range of 5.3 W to 5.5 W at 283 nm with a repetition rate of 10 kHz (i.e. $0.53-0.55$ mJ/pulse). The dye laser was tuned to excite the Q$_1(6)$ line of the $A^2 \sum^+ - X^2 \prod$ ($v^{\prime}$= 1, $v''$= 0) band. A photo-multiplier tube (PMT) with WG-305 and UG-11 filters and a reference laminar premixed flame were used to continually monitor the laser wavelength during the experiments. The 283 nm PLIF excitation beam is formed into a sheet approximately 50 mm high $\times$ 0.2 mm thick using three fused-silica, cylindrical lenses. All the lenses are coated with anti-reflective coating to maximize transmission. 

A high-speed CMOS camera (LaVision HSS 6) and an external two-stage intensifier (LaVision HS-IRO) positioned on the other side of the combustor from the OH camera are used to image the OH-PLIF fluorescence signal. Another fast UV objective lens (Cerco, $f$ = 45 mm, $f$/1.8) and a bandpass filter (300–325 nm) are fitted to the OH-PLIF camera. With an array size of 768 $\times$ 768 pixels, the camera's projected pixel resolution was 0.115 mm/pixel. OH*-chemiluminescence, OH-PLIF, and stereo-PIV images were acquired simultaneously at a sampling frequency of 10 kHz, and was phase-locked to the microphone measurements sampling data at 100 kHz. In this study, we have use only the common regions covered by all the three imaging techniques. 

%\textcolor{red}{The PIV data is then used to determine swirl number ($S$) experimentally for different conditions. The swirl number is defined as the ratio of axial flux of azimuthal momentum to the axial flux of axial momentum, and is determined according to the following expression  \citep{chigier1967experimental}},

%\begin{equation}
  %  S = \frac{G_{\phi}}{G_x R},
%\end{equation}
%where, 
%\begin{align*}
   % G_{\phi} = 2 \pi \rho \int_0^R r^2 u_y u_z dr, \enspace \textrm{and} \enspace G_x = 2 \pi \rho \int_0^R r \left(u_y^2 - \frac{1}{2}u_z^2\right) dr.
%\end{align*}
%
%Here $G_{\phi}$ and $G_x$ represent the axial flux of angular momentum and axial flux of axial momentum, respectively. $\rho$ denotes the density of the ternary mixture of air and fuel \citep{wilke1950viscosity}, while $u_y$ and $u_z$ represent the axial and tangential velocity components of the flow field. The characteristic radius for the calculation of the swirl number ($S$) is based on the diameter ($D$) of the nozzle at the inlet of the combustion chamber. $S$ ranges from 0.62 to 1.33. The swirl number ($S$) for various flow conditions are listed in \autoref{table1}. 

\section{Sampling requirement for $0-1$ chaos test}
\label{Sec2-Sampling}

\begin{figure}
%\setstretch{1.5}
  \centering
  \includegraphics[width= \textwidth]{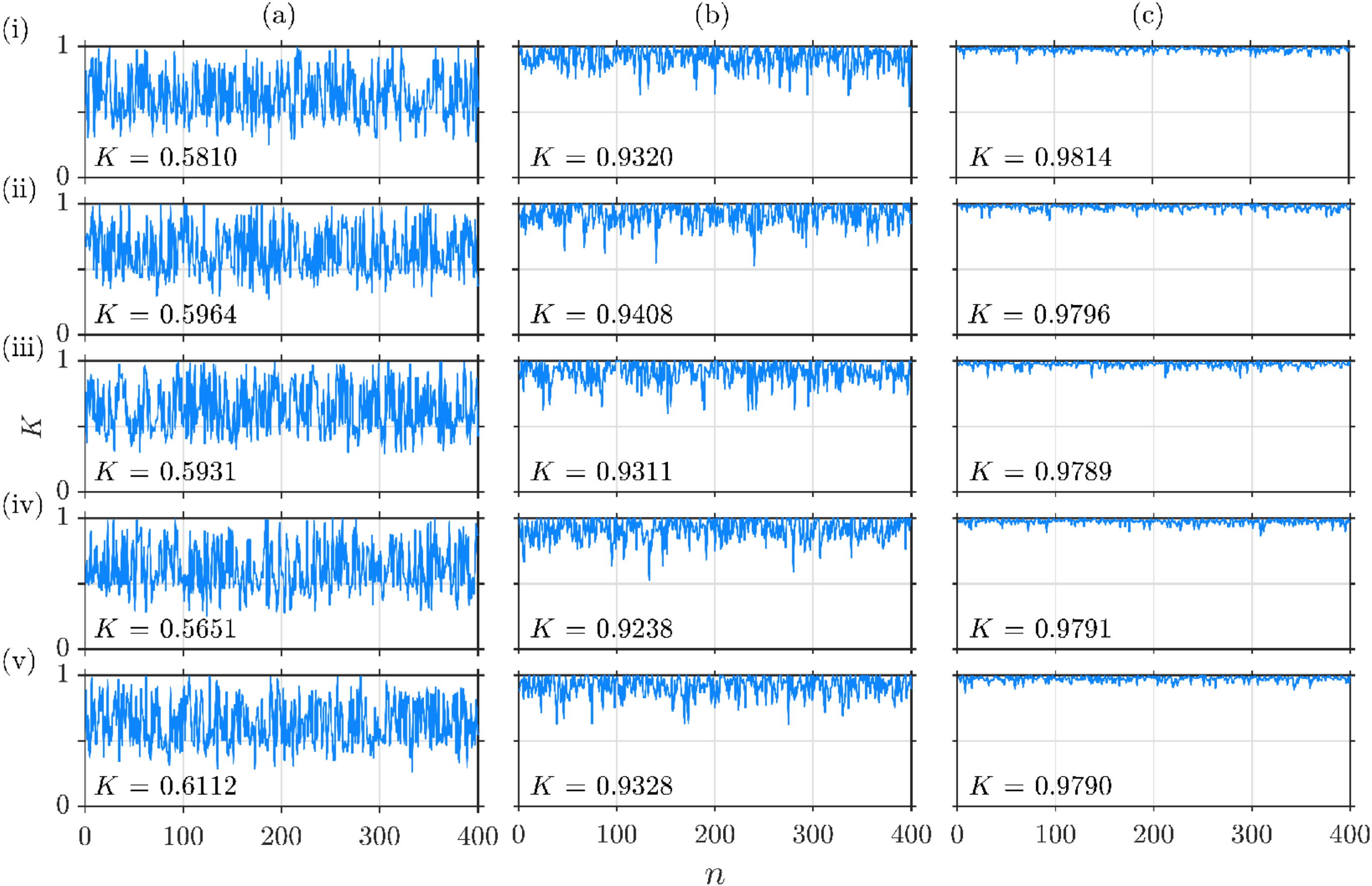}% Images in 100% size
  \caption{The variation of the asymptotic growth rate ($K$) for the signals (a) $a_1$, (b) $a_2$ and (c) heat release rate signal with (i) 10 kHz, (ii) 5 kHz, (iii) 2.5 kHz, (iv) 2 kHz, and (v) 1 kHz sampling frequency. The median value of the asymptotic growth rate is shown in each sub-figure.}
\label{figA1}
\end{figure}

We need to use the pressure data at an optimal sampling rate for performing this test \citep{armand2014modified}. The output of the 0-1 test is highly dependent on the sampling of the signal \citep{armand2014modified}. Therefore, we undersampled the original acoustic pressure signal (sampled at 100 kHz) to various levels (20 kHz, 10 kHz, 6.67 kHz, 5 kHz, 4 kHz, 3.33 kHz, 2.85 kHz and 2.5 kHz). We then tested the consistency of the chaos test using a periodic signal at the aforementioned undersampling level \citep{melosik20160}. We found that the undersampled signal having sampling rate 6.67 kHz to be consistent. Higher undersampling beyond this sampling rate leads to the destruction of key features (e.g., locations of maxima, minima, amplitude) in the oscillations of the pressure signal. Hence, we have undersampled the original signal from a sampling rate of 100 kHz to 6.67 kHz \citep{kushwaha2021dynamical}. We also confirmed our results with the recently developed novel algorithm for chaos test by \cite{toker2020simple}, called Chaos Decision Tree algorithm.

However, high speed images (PIV and OH*-chemiluminescence) are acquired at 10 kHz sampling rate which suggests that temporal coefficients of POD modes and heat release rate signals also have the sampling rate i.e., 10 kHz. We checked the signals $a_1$, $a_2$ and $\Dot{q}^{\prime}$ for different sampling rates i.e. 10 kHz, 5 kHz, 2.5 kHz. 2 kHz and 1 kHz. In figure \ref{figA1}a, we show the variation of K value for different sampling frequencies for the case of chaotic oscillations. We find that after using the correlations method, the value of K varies from 0.5 to 0.6 for $a_1$ signal which implies that it is intermittent in nature. However, by visualization of the time series of the $a_1$ signal in the manuscript, we observe aperiodic behaviour. Moreover, in case of $a_2$ and $\Dot{q}^{\prime}$, we notice that the growth rate variation near the value of 1 and the K value varies between 0.92 to 0.99. This implies that the signals $a_2$ and $\Dot{q}^{\prime}$ also show the chaotic behaviour.  

\section{Spatial results based axial velocity components}
\label{Sec3-axialcomponent EPOD}

\begin{figure}
%\setstretch{1.5}
  \centering
  \includegraphics[width=0.75\textwidth]{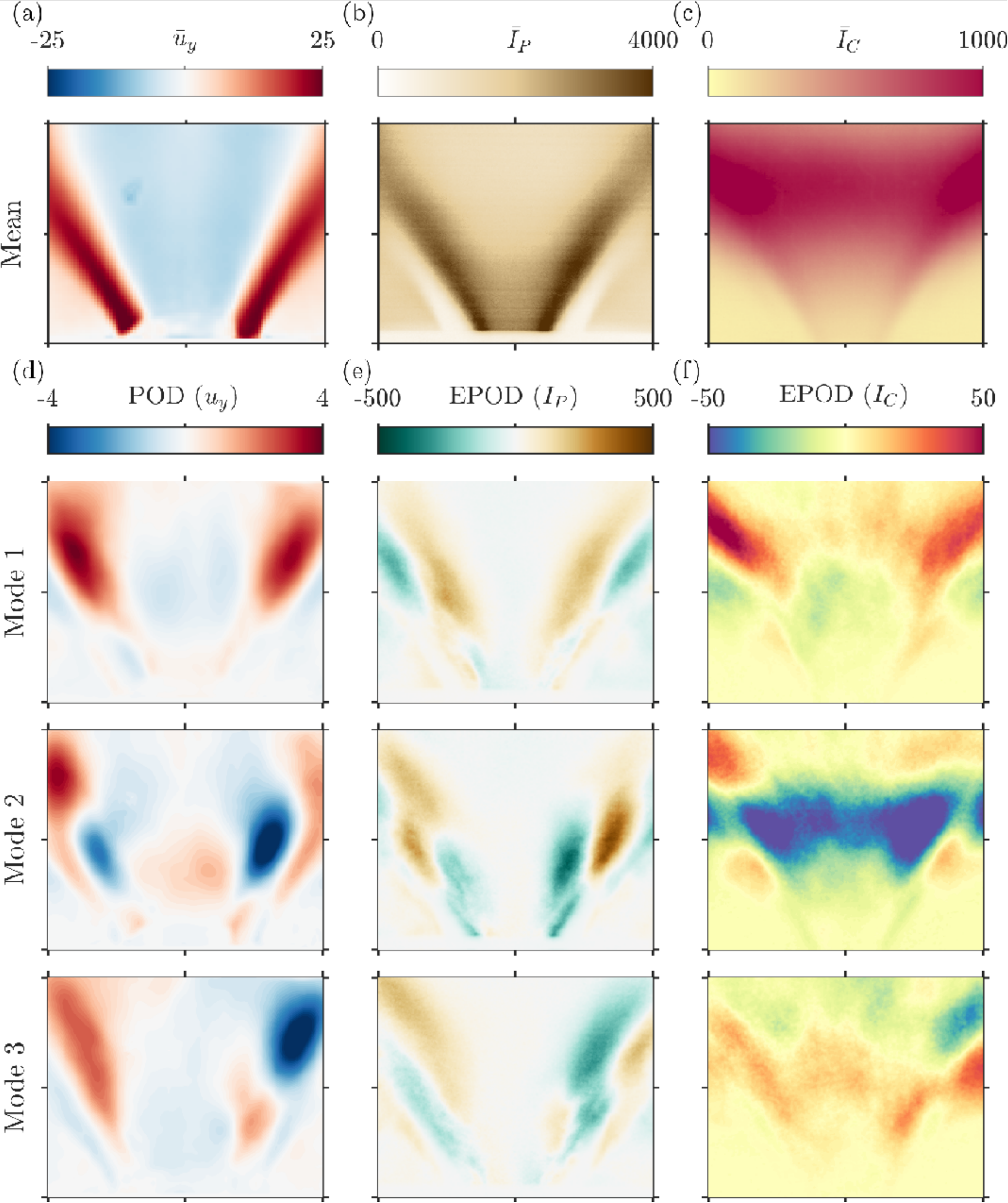}% Images in 100% size
  \caption{Flow and flame dynamics for the purely $\textrm{CH}_4$-air flame. Mean of the (a) axial velocity component ($u_y$), (b) flame brush obtained from OH-PLIF images, and (c) heat release rate fluctuations obtained from OH*-chemiluminescence. Panels (d-f) shows the first three POD modes of $u_y^\prime$ and the extended POD modes associated with $I_P^\prime$ and $I_C^\prime$. }
\label{figA2}
\end{figure}

In order to verify the differences in the EPOD modes based on transverse velocity component ($u_x$), shown in the main manuscript and the axial velocity component ($u_y$), we reconstructed the POD modes of the axial velocity for the dynamical states aforementioned, i.e., Chaotic oscillations, period-1 and period-2 limit cycle oscillations. In figure \ref{figA2}, the first row shows the mean of the axial velocity component ($\bar{u}_y$), flame brush obtained from OH-PLIF images ($\bar{I}_P$) and the heat release rate obtained from OH*-chemiluminescence images ($\bar{I}_C$). Figure \ref{figA2}a shows the high velocity along the shear layers, elongated in the radial direction toward the wall of the combustor. In addition, the inner and outer recirculation  zones have weak velocity fluctuations. The flame stabilises along the shear layers. The mean of the OH-PLIF intensity exhibits the V-shape of the flame, shown in figure \ref{figA2}b. with the increment in the flame brush thickness, most of the heat release rate occurs near the wall of the combustor (figure \ref{figA2}c) in the downstream.

Figure \ref{figA2}d shows the first POD mode of the axial component which has the high intensity structures along the shear layers, downstream of the combustor, indicating the swirling behaviour of the flow. These coherent structures grow in the axial direction and have high fluctuations near the wall of the combustor. Similarly, mode 2 and mode 3 exhibit the coherent structures along the shear layers, having difference of approximately quarter wavelength (figure \ref{figA2}d). Moreover, the first EPOD mode of the flame surfaces (figure \ref{figA2}e) exhibits small scale structures along the shear layer. Further, in mode 2 and 3, we observe the structures near the inlet of the combustor (figure \ref{figA2}e). The first EPOD mode of the heat release rate distribution shows the structures along the shear layer near the wall of the combustor (figure \ref{figA2}f). We also observe the structures in the EPOD modes of heat release rate distribution away from the inlet and near the wall of the combustor along the shear layers (figure \ref{figA2}f).

\begin{figure}
%\setstretch{1.5}
  \centering
  \includegraphics[width=0.75\textwidth]{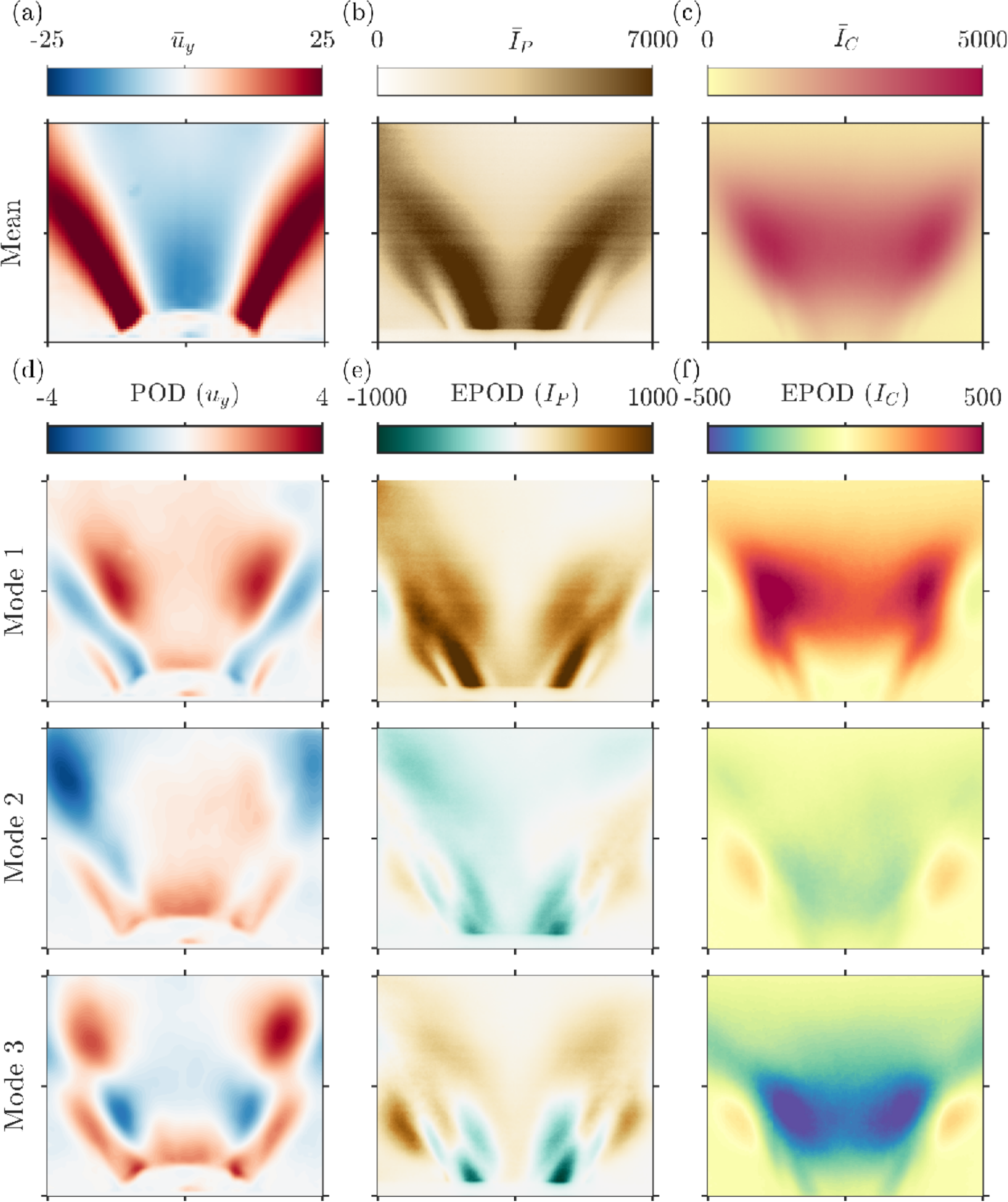}% Images in 100% size
  \caption{Flow and flame dynamics for 50\% $\textrm{H}_2$-enriched $\textrm{CH}_4$-air flame. Mean of the (a) axial velocity component ($u_y$), (b) flame brush obtained from OH-PLIF images, and (c) heat release rate fluctuations obtained from OH*-chemiluminescence. Panels (d-f) shows the first three POD modes of $u_y^\prime$ and the extended POD modes associated with $I_P^\prime$ and $I_C^\prime$.}
\label{figA3}
\end{figure}

To understand the role of coherent structures from the axial velocity during the state of thermoacoustic instability, characterized with period-1 limit cycle oscillations, we show the spatial characteristics using the time-averaged mean of the axial velocity field, the flame surface and heat release rate distribution field with the addition of hydrogen (50\%) in figure \ref{figA3}. We observe high velocity fluctuations along the shear layer, shown in figure \ref{figA3}a. Moreover, we also see the velocity fluctuations in the inner recirculation zone. Simultaneously, we also notice the M-shape flame (figure \ref{figA3}b), anchored close to the dump plane of the combustor. Most of the heat release occurs in the inner recirculation zone (IRZ) and along the shear layers (figure \ref{figA3}c).

The first three modes of the axial component of velocity are shown in figure \ref{figA3}d. The oscillations in these modes are symmetric and are in-phase. In-phase refers the same velocity fluctuations in the coherent structures. All the modes show the high energy structures at the downstream of the combustor. Moreover, we also observe high velocity fluctuations in the inner recirculation zone, indicating the existence of toroidal vortex. The first and second EPOD modes of the flame surface have the maximum intensity along the shear layers (figure \ref{figA3}e), near the inlet of the combustor. However, we observed small scale structures in the third EPOD mode (figure \ref{figA3}e). Moreover, the first and third EPOD modes of the heat release rate distribution (OH*-chemiluminescence) field show structures along the shear layer and shows the presence of toroidal vortex (figure \ref{figA3}f). However, the second EPOD mode of heat release rate distribution field shows the low intensity structures. 

\begin{figure}
%\setstretch{1.5}
  \centering
  \includegraphics[width=0.75\textwidth]{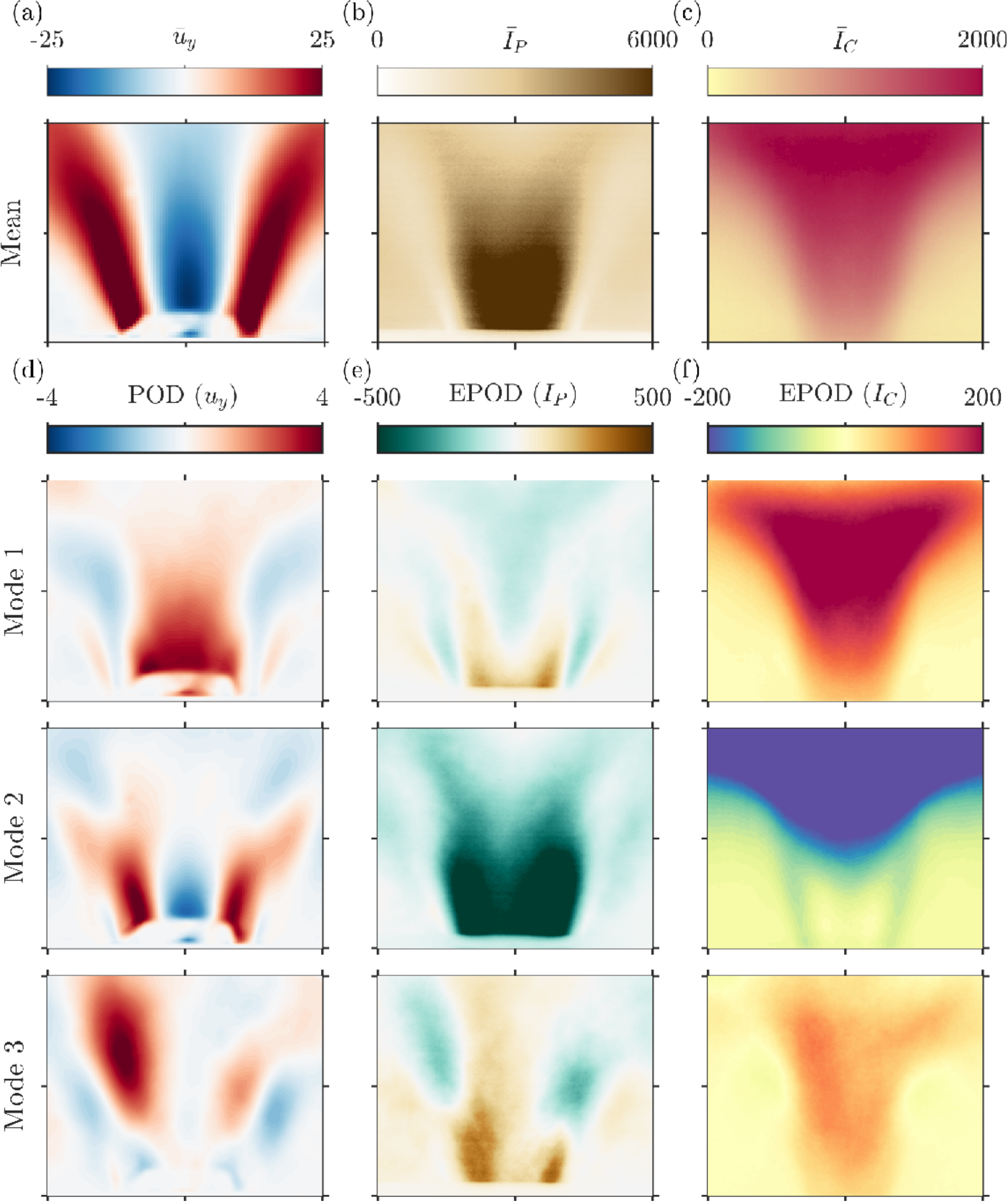}% Images in 100% size
  \caption{Flow and flame dynamics for 20\% $\textrm{H}_2$-enriched $\textrm{CH}_4$-air flame. Mean of the (a) axial velocity component ($u_y$), (b) flame brush obtained from OH-PLIF images, and (c) heat release rate fluctuations obtained from OH*-chemiluminescence. Panels (d-f) shows the first three POD modes of $u_y^\prime$ and the extended POD modes associated with $I_P^\prime$ and $I_C^\prime$.}
\label{figA4}
\end{figure}

With the addition of 20\% hydrogen at constant equivalence ratio ($\phi$) of 0.65 and thermal power ($P$) of 20 kW, we observe thermoacoustic instability with period-2 limit cycle oscillations. Figure \ref{figA4}a shows the high velocity oscillations along the shear layers and outer recirculation region (ORZ) in the mean velocity field. In figure \ref{figA4}b, the columnar shape flame having very high OH-PLIF intensity is observed near the inlet of the combustor. The most of the heat release rate occurs downstream of the combustor (figure \ref{figA4}c).

Figure \ref{figA4}d show the first three dominating POD modes of the axial velocity modes. The first mode shows the high velocity fluctuations in the IRZ, anchored with the inlet of the combustor. We also observe structures with low energy along the shear layers (figure \ref{figA4}d), revealing the helical motion in the flow. The second POD mode shows high energy structures along the shear layers near the dump plane (figure \ref{figA4}d). The inner recirculation zone also exhibits high velocity fluctuations. Moreover, the third mode exhibits small as well as large structures along the shear layers. The first and second EPOD modes of the flame surface exhibit high intensity structures along the shear layer and in the inner recirculation zone near to the inlet of the combustor (figure \ref{figA4}e). The third EPOD mode shows small small structures along the shear layers, indicating a longer flame surface (figure \ref{figA4}e). These structures correlate with the coherent structures of the axial velocity field. In figure \ref{figA4}f, the first and second EPOD of the heat release rate distribution field show that the most of the heat release rate occurs at downstream of the combustor, indicate the existence of toroidal vortex. In the third EPOD mode, we observe some heat release rate occurring in a columnar shape at the IRZ (figure \ref{figA4}f) which is similar to the third EPOD of the flame surface.

%\newpage

\bibliographystyle{jfm}
\bibliography{supplementary_material}